\documentclass[11pt]{article}
\usepackage[utf8]{inputenc}
\usepackage[T1]{fontenc}
\usepackage{lmodern}
\usepackage[margin=1in]{geometry}
\usepackage{amsmath, amssymb, amsthm}
\usepackage{graphicx}
\usepackage[colorlinks=true, allcolors=blue]{hyperref}
\usepackage[style=authoryear, maxcitenames=2, backend=biber]{biblatex}
\usepackage{tikz}
\usetikzlibrary{shapes, arrows.meta, positioning, calc, decorations.pathreplacing}
\usepackage{booktabs} % For better tables
\usepackage{tikz}
\usetikzlibrary{shapes, arrows.meta, positioning, calc, decorations.pathreplacing} % Added 'shapes' and 'arrows.meta'
\usepackage{xcolor}
\usepackage{mathrsfs}
\usepackage{mathtools}
\usepackage{csquotes}
\usepackage[
    backend=biber,     % Use Biber for processing, which is powerful and modern
    style=authoryear   % Set the citation style (e.g., (Author, Year))
]{biblatex}

\definecolor{RITorange}{RGB}{244, 121, 32} % The one true orange
\definecolor{RITbrick}{RGB}{138, 43, 26}

\usepackage[skip=10pt plus 2pt minus 2pt]{parskip} % Better paragraph spacing

\providecommand{\parenquote}[1]{\cite{#1}}

\title{No Intelligence Without Statistics: \\ The Invisible Backbone of Artificial Intelligence}
\author{Ernest Fokou\'e \\ School of Mathematics and Statistics \\ Rochester Institute of Technology \\
{\tt epfeqa@rit.edu}}
\date{\today}

\newtheorem{principle}{Principle}[section]

\begin{document}

\maketitle

\begin{abstract}
    % abstract_stat_ai.tex
%\begin{abstract}
The rapid ascent of artificial intelligence (AI) is often portrayed as a revolution born from computer science and engineering. This narrative, however, obscures a fundamental truth: the theoretical and methodological core of AI is, and has always been, statistical. This paper systematically argues that the field of statistics provides the indispensable foundation for machine learning and modern AI. We deconstruct AI into nine foundational pillars—Inference, Density Estimation, Sequential Learning, Generalization, Representation Learning, Interpretability, Causality, Optimization, and Unification—demonstrating that each is built upon century-old statistical principles. From the inferential frameworks of hypothesis testing and estimation that underpin model evaluation, to the density estimation roots of clustering and generative AI; from the time-series analysis inspiring recurrent networks to the causal models that promise true understanding, we trace an unbroken statistical lineage. 
While celebrating the computational engines that power modern AI, we contend that statistics provides the \textit{brain}—the theoretical frameworks, uncertainty quantification, and inferential goals—while computer science provides the \textit{brawn}—the scalable algorithms and hardware. Recognizing this statistical backbone is not merely an academic exercise, but a necessary step for developing more robust, interpretable, and trustworthy intelligent systems. We issue a call to action for education, research, and practice to re-embrace this statistical foundation. Ignoring these roots risks building a fragile future; embracing them is the path to truly intelligent machines. \textbf{There is no machine learning without statistical learning; no artificial intelligence without statistical thought.}
%\end{abstract}
\end{abstract}

%\section{Introduction: The Statistical Nature of Learning}
\section{From the Grammar of Science to the Engine of AI: The Statistical Imperative}
\noindent If we wish to build intelligent systems, we must first ask: what is intelligence, but the capacity to find meaningful patterns and make reliable predictions in a complex, uncertain world? This is not a new problem; it is the fundamental problem of science itself. Karl Pearson, in his seminal work \textit{The Grammar of Science} \autocite{pearson1930}, asserted that "the whole field of scientific investigation is co-extensive with the applications of statistics." In this light, statistics is not merely a branch of mathematics; it is the very grammar—the underlying logical structure—for reasoning from empirical evidence.

This pursuit is rooted in the human condition. We are, by nature, pattern-seeking creatures, instinctively calculating central tendencies and assessing variances in our environment. Our philosophical traditions have long grappled with the chasm between observation and truth. Plato's allegory of the cave reminds us that we see but shadows of a deeper reality—shadows that statistics treats as data, from which we must infer the nature of the true forms. William of Ockham's razor, a principle of parsimony, finds its formal expression in statistical regularization and model selection, penalizing unnecessary complexity. Karl Popper's doctrine of falsification is operationalized through statistical hypothesis testing, where we can never confirm a model, only fail to reject it \autocite{fisher1935}.

The mathematical machinery for this endeavor was forged in the fires of uncertainty. The correspondence between Blaise Pascal and Pierre de Fermat gave birth to probability theory. Thomas Bayes and Pierre-Simon Laplace laid the groundwork for inductive inference, providing a formal mechanism for updating beliefs in the light of new data \autocite{laplace1951}. This lineage culminated in the work of Francis Galton, Karl Pearson, and R.A. Fisher, who transformed these ideas into the modern discipline of statistics, with its core concepts of correlation, regression, and maximum likelihood estimation \autocite{pearson1901, fisher1922}. Eugene Wigner marveled at the "unreasonable effectiveness of mathematics in the natural sciences" \autocite{wigner1960}; we contend that it is the reasonable effectiveness of statistics that is truly indispensable, for it provides the framework to manage the uncertainty that Wigner's idealized mathematics often ignores.

The 20th century witnessed a symbiotic convergence. The theoretical foundations of computation, laid by Alan Turing and John von Neumann, provided the physical engine. Yet, it is crucial to recognize that this engine was, from its inception, imbued with a statistical soul. Von Neumann's architecture was designed for deterministic, procedural calculation. But the problems of the real world—the "shadows in the cave"—are not deterministic. They are noisy, incomplete, and probabilistic. The great leap was the realization that this rigid computational engine could be harnessed to execute flexible, fault-tolerant statistical learning processes. Claude Shannon's information theory \autocite{shannon1948} provided the currency, and figures like Turing himself, with his Bayesian approach to code-breaking, exemplified the fusion. The Heisenberg Uncertainty Principle in physics finds its parallel in the bias-variance trade-off in statistics: a fundamental, inescapable tension between accuracy and precision in any measurement or model.

It is against this rich historical and philosophical tapestry that the ascent of artificial intelligence must be viewed. The common narrative celebrates the computational horsepower—the faster processors and larger memories—while ignoring the intellectual software that gives them purpose. This paper aims to correct this narrative. We posit that \textbf{the theoretical and methodological core of AI is, and has always been, statistical}. The excitement around deep learning and generative AI has created a form of historical amnesia, where the extensive statistical lineage of these methods is often ignored or forgotten. As \textcite{lecun2015} themselves acknowledge, deep learning's optimization, generalization, and inference frameworks are thoroughly statistical. \textcite{pearl2009} laments the lack of causal reasoning in modern ML, a field built entirely upon statistical foundations.

This paper is therefore a reclamation project. It is structured as a journey through the nine fundamental pillars of statistics that not only support but constitute the very essence of AI. We will demonstrate that from its theoretical roots to its modern algorithms, artificial intelligence is applied statistics at scale. 
This is not to diminish the monumental achievements of computer science in creating the scalable algorithms and hardware that realize these statistical principles. Rather, it is to clarify their respective roles: statistics provides the what and why of learning, while computer science provides the how.
Our argument is not one of mere precedence, but of essential identity. To build the future of AI on a stable foundation, we must first acknowledge the ground on which it stands.

% New subsection for the introduction_stat_ai.tex
\subsection{The Human Engine: From Statistical Calculation to Computational Learning}

Long before silicon, there was flesh and blood. The first "computers" were not machines, but humans—often teams of highly skilled women, such as those at the Indian Statistical Institute under Mahalanobis and at the University of Pennsylvania's Moore School—who performed the monumental calculations required for multivariate analysis, ballistics tables, and early economic modeling. They were executing statistical algorithms by hand. This historical fact underscores a profound truth: the \textit{statistical questions came first}; the electronic computer was invented as a tool to solve them at scale. 

The design of these studies—\textbf{how} to collect data to best answer a specific question—is the domain of experimental design, a cornerstone of statistical science formalized by R.A. Fisher. This principle is experiencing a renaissance in modern AI through paradigms like \textbf{active learning}, where the model itself statistically guides the data acquisition process to maximize information gain. The cycle is complete: statistical thinking now guides the computational engine on what data to compute on.

% New subsection for pillar1_inference_stat_ai.tex
\subsubsection{The Generalized Linear Model (GLM) as a Proto-Neural Network}

A profound, yet often overlooked, statistical precursor to deep learning is the Generalized Linear Model (GLM) framework \parencite{nelder1972}. A GLM makes a prediction via:
\[
\mathbb{E}[Y] = g^{-1}(\mathbf{w}^\top \mathbf{x})
\]
where $g$ is the link function. This is structurally identical to a single-layer neural network with activation function $g^{-1}$.

\begin{itemize}
    \item \textbf{Logistic Regression} (a GLM for classification) uses a sigmoid link, precisely the activation function used in the output layer of a binary classification network.
    \item \textbf{Poisson Regression} uses an exponential link, analogous to an output layer for count data.
\end{itemize}

A Deep Neural Network (DNN) can be viewed as a \textbf{deep, hierarchical GLM}. Instead of a single linear combination $\mathbf{w}^\top \mathbf{x}$, a DNN learns a cascade of latent representations $\mathbf{h}^{(l)} = \sigma(\mathbf{W}^{(l)}\mathbf{h}^{(l-1)})$, culminating in a final GLM-like layer. The cross-entropy loss used to train these networks is the same negative log-likelihood derived for logistic regression and other GLMs. The backpropagation algorithm is, therefore, the computational method for performing MLE in this complex, composable statistical architecture. The deep learning revolution did not abandon statistics; it provided the machinery to estimate parameters for \textit{deep} generalized linear models.

%\section{Pillar I: Inference Under Uncertainty}
% sections/pillar1_inference.tex
\section{Inference Under Uncertainty (Pillar I)}
\label{sec:inference}

The very concept of learning from data is an exercise in navigating uncertainty. It is the quintessential statistical problem. The entire framework of modern machine learning (ML)—from estimating model parameters to evaluating their performance—is built upon the bedrock of statistical inference. This pillar argues that \textbf{the core operations of ML are not merely analogous to statistical inference; they are direct implementations of its formalisms}.

\begin{principle}[The Primacy of Inference]
All learning is inference. Training a model is parameter estimation; evaluating it is hypothesis testing; and quantifying its uncertainty is interval estimation.
\end{principle}

\subsection{Parameter Estimation: From Likelihood to Loss Functions}

The most direct lineage from statistics to ML is found in parameter estimation. The concept of finding the ``best'' parameters $\boldsymbol{\theta}$ for a model given observed data $\mathscr{D}_n = \{(\mathbf{x}_i, y_i)\}_{i=1}^n$ is a statistical one.

\subsubsection{Maximum Likelihood Estimation (MLE)}

The principle of MLE, formalized by \textcite{fisher1922}, is the statistical engine at the heart of most supervised learning. The goal is to find the parameter vector $\boldsymbol{\theta}$ that maximizes the likelihood of the observed data $\mathscr{D}_n$:
\[
\hat{\boldsymbol{\theta}}_{\text{MLE}} = \arg\max_{\boldsymbol{\theta}} \, p(\mathscr{D}_n \mid \boldsymbol{\theta})
\]

This foundational statistical concept manifests in ML as the minimization of a loss function. For example:
\begin{itemize}
    \item \textbf{Linear Regression}: Minimizing Mean Squared Error (MSE) is equivalent to MLE under a Gaussian noise assumption \parencite{hastie2009}.
    \item \textbf{Logistic Regression}: Minimizing log-loss (cross-entropy) is equivalent to MLE for a Bernoulli distribution \parencite{hastie2009}.
    \item \textbf{Deep Learning:} The cross-entropy loss used in classification tasks is precisely the negative log-likelihood of the data given the model parameters.
\end{itemize}

The backpropagation algorithm \parencite{rumelhart1986} and stochastic gradient descent (SGD) are \textbf{computational techniques} for solving this statistical optimization problem, whose theoretical foundations lie in the Robbins-Monro algorithm \parencite{robbins1951}. Here, computer science provides the \textit{scalable engine} for the statistical \textit{learning objective}.

\subsubsection{Bayesian Inference}
While MLE provides point estimates, Bayesian inference provides a full posterior distribution over parameters, naturally quantifying uncertainty. The Bayesian paradigm \parencite{gelman2013}:
\[
P(\boldsymbol{\theta} \mid \mathscr{D}_n) = \frac{P(\mathscr{D}_n \mid \boldsymbol{\theta}) \, P(\boldsymbol{\theta})}{P(\mathscr{D}_n)}
\]
has been adopted wholeheartedly in ML:
\begin{itemize}
    \item \textbf{Bayesian Neural Networks (BNNs)} \parencite{mackay1992} place distributions over weights, transforming point estimates into predictive distributions.
    \item \textbf{Gaussian Processes} \parencite{rasmussen2006} are a non-parametric Bayesian approach for regression, providing uncertainty estimates directly.
    \item Variational Inference (VI) \parencite{jordan1999} and Markov Chain Monte Carlo (MCMC) methods \parencite{neal1993} are statistical techniques developed for intractable Bayesian calculations and are now central to probabilistic ML.
\end{itemize}

\subsection{Model Evaluation as Hypothesis Testing}

The process of evaluating and comparing models is a direct application of statistical hypothesis testing.

\subsubsection{The Null Hypothesis in ML}

The question ``Does my model perform better than random?'' is a hypothesis test. Techniques like \textbf{permutation tests} \parencite{efron1994} are used rigorously to compute p-values for model accuracy, testing the null hypothesis that the model has no predictive power.

\subsubsection{A/B Testing and Reinforcement Learning}

The entire field of A/B testing \parencite{kohavi2020} is applied statistical hypothesis testing. This framework is crucial in industry for evaluating ML systems and is the foundation for \textbf{multi-armed bandit} algorithms \parencite{lattimore2020} and \textbf{policy evaluation} in Reinforcement Learning (RL). An RL agent exploring an environment is fundamentally conducting sequential hypothesis tests to distinguish profitable actions from non-profitable ones.

\subsection{Uncertainty Quantification: From Confidence Intervals to Predictive Distributions}

A model's prediction is meaningless without a measure of its uncertainty. This concept is purely statistical.

\subsubsection{Frequentist Uncertainty}

In traditional statistics, confidence intervals express the uncertainty in an parameter estimate. In ML, this translates to:
\begin{itemize}
    \item \textbf{Confidence Intervals} for predictions, commonly derived from bootstrapping \parencite{efron1994} or analytical methods.
    \item \textbf{Calibration} \parencite{niculescu-mizil2005}: The process of ensuring a classifier's predicted probability (e.g., 0.8 for class ``cat'') matches the empirical frequency (e.g., 80\% of instances with score 0.8 are indeed cats). A well-calibrated model is one whose uncertainty estimates are statistically faithful.
\end{itemize}

\subsubsection{Bayesian Uncertainty}

As mentioned, Bayesian methods provide a natural framework for uncertainty. The \textbf{predictive posterior distribution}
\[
P(y^* \mid \mathbf{x}^*, \mathscr{D}_n) = \int P(y^* \mid \mathbf{x}^*, \boldsymbol{\theta}) \, P(\boldsymbol{\theta} \mid \mathscr{D}_n) \, d\boldsymbol{\theta}
\]
is the complete Bayesian answer, providing a distribution over possible outcomes, not just a single point. This is the gold standard for uncertainty quantification in ML.

\subsection{Conclusion of Pillar I}

The machinery of statistical inference is not a tool occasionally used by ML; it is the very substrate on which ML is built. The journey from a dataset $\mathscr{D}_n$ to a trained, evaluated, and deployed model is a continuous thread of statistical reasoning:
\begin{enumerate}
    \item \textbf{Estimation} (MLE/Bayes) $\rightarrow$ Defines the learning objective.
    \item \textbf{Testing} (Hypothesis Tests) $\rightarrow$ Validates the model's utility.
    \item \textbf{Uncertainty Quantification} (Intervals/Distributions) $\rightarrow$ Informs trustworthy deployment.
\end{enumerate}

To ignore this is to practice ancient primordial rudimentary alchemy. To embrace it is to practice science. \textbf{The entire architecture of artificial intelligence is built upon this statistical foundation.}

%\section{Pillar II: Density Estimation and Novelty Detection}
% sections/pillar2_density.tex
\section{Density Estimation and Novelty Detection (Pillar II)}
\label{sec:density}

If the first pillar established that learning is inference, the second asserts that \textbf{to learn is to model probability distributions}. The fundamental statistical problem of estimating the underlying probability density function $p(\mathbf{x})$ from observed data $\mathscr{D}_n = \{\mathbf{x}_1, \ldots, \mathbf{x}_n\}$ is the engine behind some of the most transformative subfields of AI: clustering, anomaly detection, and generative modeling. This pillar reveals that these advanced capabilities are not novel inventions but sophisticated scaling of core statistical concepts.

\begin{principle}[The Density Estimation Principle]
Unsupervised learning is density estimation. Clustering, novelty detection, and generative modeling are all manifestations of the statistical problem of estimating $p(\mathbf{x})$ or $p(\mathbf{x}, y)$.
\end{principle}

\subsection{Clustering: Modeling Mixtures of Distributions}

The seemingly modern task of partitioning data into groups has its roots in one of the oldest problems in statistics: modeling population heterogeneity.

\subsubsection{Mixture Models and the EM Algorithm}

The statistical formalization of clustering is the mixture model, where we assume data is generated from a convex combination of $K$ underlying probability distributions (e.g., Gaussians) \parencite{mclachlan2004}:
\[
p(\mathbf{x}) = \sum_{k=1}^K \pi_k \, p(\mathbf{x} \mid \boldsymbol{\theta}_k)
\]
where $\pi_k$ are the mixing coefficients. Learning the parameters $\boldsymbol{\theta}_k$ and the latent cluster assignments $z$ for each data point is a classic statistical problem solved by the Expectation-Maximization (EM) algorithm \parencite{dempster1977}.

This is not merely a historical note; it is the foundational theory. The ubiquitous \textbf{K-Means} algorithm \parencite{lloyd1982} is a specific case of EM applied to Gaussian mixture models with isotropic covariances and hard assignments.

\subsubsection{Non-Parametric Clustering}

Beyond mixtures, other clustering paradigms are also statistically grounded:
\begin{itemize}
    \item \textbf{Hierarchical Clustering} leverages statistical measures of distance and linkage.
    \item \textbf{DBSCAN} \parencite{ester1996} is based on the statistical concept of kernel density estimation (KDE), identifying clusters as high-density regions separated by low-density regions.
\end{itemize}

\subsection{Novelty Detection: Density Estimation in the Tails}

The identification of rare, unusual, or anomalous events is a quintessential statistical task focused on the tails of a distribution $p(\mathbf{x})$.

\subsubsection{The Statistical Basis of Anomaly Detection}

Anomaly detection algorithms are directly derived from statistical principles:
\begin{itemize}
    \item \textbf{Parametric Methods:} Assume a parametric form for $p(\mathbf{x})$ (e.g., Gaussian) and flag points with low probability density \parencite{aggarajan2017}.
    \item \textbf{Non-Parametric Methods:} Methods like \textbf{Histogram-Based Outlier Detection} and \textbf{Kernel Density Estimation (KDE)} \parencite{parzen1962} estimate $p(\mathbf{x})$ non-parametrically and apply a threshold.
    \item \textbf{Depth-Based Methods:} Concepts like \textbf{Mahalanobis Distance} \parencite{mahalanobis1936} measure how many standard deviations a point is from the mean of a distribution, accounting for covariance.
\end{itemize}

\subsubsection{Modern incarnations: From Statistical Roots to AI Algorithms}

Modern ML anomaly detection techniques are direct descendants of these statistical ideas:
\begin{itemize}
    \item \textbf{Isolation Forest} \parencite{liu2008} is fundamentally a non-parametric method that isolates anomalies based on their propensity to be separated with fewer random partitions—a concept related to their low probability mass.
    \item \textbf{One-Class Support Vector Machines (SVMs)} \parencite{scholkopf2001} learn a tight boundary around the data, which can be interpreted as estimating a density level set.
\end{itemize}

\subsection{Generative Modeling: The Pinnacle of Density Estimation}

The recent revolution in generative AI is, at its core, a dramatic scaling of the statistical problem of density estimation.

\subsubsection{Generative Adversarial Networks (GANs)}

The GAN framework \parencite{goodfellow2014} poses a brilliant game-theoretic approach to density estimation. The generator learns to transform a simple noise distribution $p(\mathbf{z})$ into a complex data distribution $p_{\text{model}}(\mathbf{x})$ that approximates $p_{\text{data}}(\mathbf{x})$. This is a complex, implicit method of learning to sample from an estimated density.

\subsubsection{Variational Autoencoders (VAEs)}

VAEs \parencite{kingma2013} provide a more explicitly statistical approach. They combine the neural network function approximation of an autoencoder with the Bayesian probabilistic framework of \textbf{variational inference} \parencite{jordan1999}. The VAE learns a latent variable model $p(\mathbf{x} \mid \mathbf{z})$ and approximates the posterior $p(\mathbf{z} \mid \mathbf{x})$, directly tackling the intractable integrals of Bayesian statistics with deep learning.

\subsubsection{Normalizing Flows}

This family of models \parencite{papamakarios2021} performs exact density estimation by learning invertible, differentiable transformations between a simple base distribution (e.g., Gaussian) and the complex data distribution. This allows for both efficient sampling and exact calculation of $p(\mathbf{x})$, representing a pure and powerful fusion of deep learning and statistical density estimation.

\subsection{Conclusion of Pillar II}

The path from a simple histogram to a state-of-the-art generative model is a continuous spectrum of solving the same core statistical problem: \textit{What does the probability distribution of this data look like?}

\begin{itemize}
    \item \textbf{Clustering} asks: ``Is the data a mixture of simpler distributions?''
    \item \textbf{Anomaly Detection} asks: ``Does this new point lie in a low-probability region of the estimated distribution?''
    \item \textbf{Generative Modeling} asks: ``Can I so perfectly estimate $p(\mathbf{x})$ that I can sample new, realistic data from it?''
\end{itemize}

The algorithms have evolved from simple equations to deep neural networks, but the statistical soul remains unchanged. \textbf{Statistics is the grammar, the logic, and the soul of machine intelligence} in the unsupervised realm. The statistical footprints in clustering, anomaly detection, and generative AI are not merely permeating the landscape---they \textit{are} the landscape.

%\section{Pillar III: Time Series and Sequential Learning}
% sections/pillar3_timeseries.tex
\section{Time Series, Sequential Learning, and the Architecture of Memory (Pillar III)}
\label{sec:timeseries}

The world is not a collection of independent data points; it is a sequence of interdependent events. The statistical challenge of modeling temporal dependence—where the value at time $t$ depends on the history of values up to $t-1$—is the cornerstone of forecasting, signal processing, and sequential decision-making. This pillar demonstrates that the sophisticated neural architectures developed for sequences are direct conceptual and formal descendants of classical time series models, scaled by depth and data. \textbf{There is no sequential intelligence without statistical time series analysis}.

\begin{principle}[The Sequential Principle]
Modeling sequences requires explicitly representing dependency through time. The statistical formulation of this problem—from autoregressive models to state-space representations—provides the essential blueprint for all sequential AI.
\end{principle}

\subsection{Autoregression: The Statistical Engine of Prediction}

The foundational statistical idea for sequential prediction is that a future value can be forecasted by a linear combination of its past values.

\subsubsection{The ARIMA Framework}

The Autoregressive Integrated Moving Average (ARIMA) model \parencite{box1976} is a pinnacle of statistical time series analysis. Its components form the conceptual DNA of modern sequence models:
\begin{itemize}
    \item \textbf{Autoregressive (AR)} Component: $y_t = c + \sum_{i=1}^p \phi_i y_{t-i} + \varepsilon_t$ \\The value at time $t$ is a weighted sum of the previous $p$ values. This is a linear form of memory.
    \item \textbf{Moving Average (MA)} Component: $y_t = \mu + \varepsilon_t + \sum_{i=1}^q \theta_i \varepsilon_{t-i}$ \\The model incorporates the memory of past forecast errors, a form of adaptive correction.
\end{itemize}
The AR component's principle—\textit{explain the present with the recent past}—is the core insight that scales into nonlinear deep learning architectures.

\subsubsection{From Linear AR to Nonlinear Neural Autoregression}

The leap from statistical to neural autoregression is one of function approximation:
\begin{itemize}
    \item A linear AR($p$) model: $y_t = f(y_{t-1}, ..., y_{t-p}) = \boldsymbol{\phi}^\top \mathbf{y}_{t-p:t-1}$
    \item A neural AR model: $y_t = f_{\boldsymbol{\theta}}(y_{t-1}, ..., y_{t-p})$, where $f_{\boldsymbol{\theta}}$ is a neural network. \\This is the architecture of a \textbf{Time-Lagged Feedforward Network} \parencite{weigend1993}, the direct neural incarnation of the AR concept.
\end{itemize}

\subsection{State-Space Models: The Statistical Theory of Hidden States}

A more powerful statistical framework posits that observed data is generated from a hidden (latent) state that evolves over time.

\subsubsection{Kalman Filter and Linear Gaussian State-Space Models}

The Kalman Filter \parencite{kalman1960} provides an elegant recursive solution to a continuous state-space model:
\begin{align*}
\text{State Transition:} \quad & \mathbf{z}_t = \mathbf{A}\mathbf{z}_{t-1} + \boldsymbol{\epsilon}_t \quad \boldsymbol{\epsilon}_t \sim \mathcal{N}(0, \mathbf{Q}) \\
\text{Emission:} \quad & \mathbf{x}_t = \mathbf{C}\mathbf{z}_t + \boldsymbol{\delta}_t \quad \boldsymbol{\delta}_t \sim \mathcal{N}(0, \mathbf{R})
\end{align*}
This provides the optimal Bayesian update for estimating the hidden state $\mathbf{z}_t$ given observations $\mathbf{x}_{1:t}$.

\subsubsection{The Nonlinear Leap: From Kalman to Recurrent Networks}

Recurrent Neural Networks (RNNs) \parencite{rumelhart1986} are the nonlinear, learnable generalization of state-space models:
\begin{align*}
\text{Hidden State Update:} \quad & \mathbf{h}_t = \sigma(\mathbf{W}_h \mathbf{h}_{t-1} + \mathbf{W}_x \mathbf{x}_t + \mathbf{b}) \\
\text{Output:} \quad & \mathbf{y}_t = \text{softmax}(\mathbf{W}_y \mathbf{h}_t + \mathbf{b}_y)
\end{align*}
Here, the hidden state $\mathbf{h}_t$ is the analog of the statistical latent state $\mathbf{z}_t$. The function $\sigma$ (e.g., tanh) introduces nonlinearity, and the matrices $\mathbf{W}$ are learned from data instead of being specified by a modeler.

The Long Short-Term Memory (LSTM) \parencite{hochreiter1997} and Gated Recurrent Unit (GRU) \parencite{cho2014} architectures can be viewed as sophisticated, learned algorithms for managing this hidden state, solving the statistical problem of vanishing gradients, which is the practical equivalent of maintaining memory over long sequences.

\subsection{Attention and Transformers: The Statistical Mechanics of Relevance}

The latest revolution in sequence modeling emerged from a statistical idea: not all past information is equally relevant for predicting the present.

\subsubsection{Weighted Regression as a Proto-Attention Mechanism}

The core idea of attention is to compute a context vector as a \textit{weighted average} of previous hidden states. This is a direct statistical concept. \textbf{Nadaraya-Watson kernel regression} \parencite{nadaraya1964, watson1964} estimates a function as a weighted average of training examples:
\[
E[y | x] = \frac{\sum_i K(x, x_i) y_i}{\sum_i K(x, x_i)}
\]
where the kernel $K$ measures similarity. This is precisely the mathematical form of attention: the output is a weighted sum of values, where the weights (attention scores) are a function of the similarity between a query and a set of keys.

\subsubsection{The Transformer Architecture as Statistical Weighting Scaled to Infinity}

The Transformer \parencite{vaswani2017} is the ultimate manifestation of this statistical idea. Its self-attention mechanism:
\[
\text{Attention}(\mathbf{Q}, \mathbf{K}, \mathbf{V}) = \text{softmax}\left(\frac{\mathbf{Q}\mathbf{K}^\top}{\sqrt{d_k}}\right)\mathbf{V}
\]
is a learnable, parameterized version of kernel smoothing. The query, key, and value matrices allow the model to learn which aspects of the sequence history are most relevant for a given context, performing a form of non-parametric estimation at a massive scale. The statistical footprint is unmistakable: \textbf{attention is data-driven kernel regression over sequences}.

\subsection{Conclusion of Pillar III}

The evolution from statistical time series models to modern deep sequential AI is a story of increasing flexibility and scale, not a change in fundamental purpose:
\begin{enumerate}
    \item \textbf{Autoregression (AR)} provided the principle: The past predicts the future.
    \item \textbf{State-Space Models (Kalman)} provided the architecture: Maintain a hidden state that evolves.
    \item \textbf{Recurrent Networks (RNNs/LSTMs)} provided the learnable, nonlinear implementation.
    \item \textbf{Attention/Transformers} provided the optimal weighting: Dynamically determine which parts of the past are most relevant.
\end{enumerate}

Each layer of intelligence in sequence modeling rests upon this bedrock of statistical principles. The journey from the Kalman filter to the LSTM to the Transformer is one of brilliant engineering built upon a timeless statistical foundation. The statistical footprints in time series analysis and sequential AI are not merely permeating the landscape—they \textit{are} the landscape.

%\section{Pillar IV: Generalization}
% sections/pillar4_generalization.tex

\section{Generalization, Assessment, and the Science of Trustworthy AI (Pillar IV)}
\label{sec:generalization}

A model that performs perfectly on its training data $\mathscr{D}_n$ is a parlor trick. A model that reliably performs well on new, unseen data is intelligence. The chasm between these two is the fundamental problem of generalization, and bridging it is the primary purpose of statistical theory in machine learning. This pillar demonstrates that the entire modern apparatus for evaluating, validating, and trusting AI systems—from cross-validation to performance bounds—is built upon a foundation of statistical ideas developed throughout the 20th century. \textbf{There is no trustworthy AI without statistical generalization}.

\begin{principle}[The Generalization Principle]
The ultimate goal of learning is performance on the population, not the sample. All techniques for estimating and ensuring this performance are statistical in nature.
\end{principle}

\subsection{Resampling: The Empirical Bridge to the Population}

Since we cannot access the true data-generating distribution, we must estimate the generalization error using the sample we have. This is the classic statistical problem of estimation, solved by ingenious resampling methods.

\subsubsection{Cross-Validation: The Workhorse of Modern ML}

Cross-validation (CV) is the direct statistical solution to the problem of estimating out-of-sample error using only in-sample data.

\begin{itemize}
    \item \textbf{The Core Idea:} The data is partitioned, a model is trained on a subset and validated on the held-out portion. This process is repeated to simulate performance on unseen data.
    \item \textbf{Statistical Lineage:} While the basic idea of data splitting is ancient, its formalization into $k$-fold CV is a cornerstone of modern statistical practice \parencite{stone1974, geisser1975}. It is an empirical method for estimating the expected prediction error, $E[L(Y, \hat{f}(X))]$.
    \item \textbf{Ubiquity in AI:} $k$-fold CV is the standard method for hyperparameter tuning and model selection across all of ML. More complex variants like \textbf{nested cross-validation} provide nearly unbiased estimates of generalization error \parencite{varma2006}.
\end{itemize}

\subsubsection{The Bootstrap: Efron's Masterpiece}

If CV estimates generalization error, the bootstrap \parencite{efron1979} quantifies its uncertainty. It is a quintessentially statistical idea that has become indispensable for understanding model stability.

\begin{itemize}
    \item \textbf{The Core Idea:} By creating many simulated datasets through resampling with replacement, we can approximate the sampling distribution of almost any statistic (e.g., test accuracy, model parameters) without stringent parametric assumptions.
    \item \textbf{The Connection to Bagging:} Breiman's Bagging (Bootstrap Aggregating) \parencite{breiman1996} is a direct application of the bootstrap. By training multiple models on bootstrap samples and aggregating their predictions, Bagging reduces variance and improves generalization—a statistical ensemble method powered by a statistical resampling technique.
    \item \textbf{Uncertainty for Black Boxes:} The bootstrap is widely used to compute confidence intervals for the predictions of complex models like random forests and neural networks, providing a crucial layer of statistical trustworthiness.
\end{itemize}

\subsection{Learning Theory: The Mathematical Guarantees}

Beyond empirical methods, statistics provides the theoretical machinery to derive mathematical bounds on generalization error, explaining \textit{why} learning is even possible.

\subsubsection{The Vapnik-Chervonenkis (VC) Theory}

The VC theory \parencite{vapnik1971} provides a probabilistic bound on the difference between empirical risk (training error) and expected risk (generalization error). For a classifier with VC dimension $d$, with probability at least $1-\eta$:
\[
R(\hat{f}) \leq R_n(\hat{f}) + \Phi(n, d, \eta)
\]
where $\Phi$ is a function that grows with model complexity $d$ and decreases with sample size $n$. This statistical result formally characterizes the \textbf{bias-variance tradeoff}, the central dogma of statistical learning.

\subsubsection{Concentration Inequalities: The Statistical Engine}

The VC bound, and many others like it, are powered by profound statistical results that quantify how random variables concentrate around their mean.
\begin{itemize}
    \item \textbf{Hoeffding's Inequality} bounds the deviation of a sample average from its expectation.
    \item \textbf{McDiarmid's Inequality} generalizes this to functions of independent random variables.
\end{itemize}
These are not mere mathematical curiosities; they are the bedrock of generalization theory. Every PAC (Probably Approximately Correct) learning bound \parenquote{valiant1984} is ultimately derived from such statistical inequalities.

\subsection{Calibration: When a Probability is Really a Probability}

A model's ability to generalize is not just about getting the answer right, but also about knowing when it is uncertain. This is the statistical problem of calibration.

\subsubsection{The Statistical Goal of Calibration}

A model is perfectly calibrated if its predicted probability reflects the empirical frequency. For example, of all instances where the model predicts a probability of 0.8, exactly 80\% should belong to the positive class. This is a direct statistical statement about the agreement between a subjective probability (the prediction) and an objective frequency \parencite{dawid1982}.

\subsubsection{Modern Metrics and Methods}

The assessment and improvement of calibration are purely statistical endeavors:
\begin{itemize}
    \item \textbf{Evaluation:} Metrics like the \textbf{Expected Calibration Error (ECE)} and \textbf{Brier Score} \parencite{brier1950} are used to measure miscalibration.
    \item \textbf{Techniques:} Methods like \textbf{Platt Scaling} \parencite{platt1999} and \textbf{Isotonic Regression} are post-processing techniques designed to statistically calibrate the outputs of a trained model, ensuring its uncertainty estimates are trustworthy.
\end{itemize}

\subsection{Conclusion of Pillar IV}

The journey from a trained model to a deployed, trustworthy AI system is a journey through statistical inference:
\begin{enumerate}
    \item \textbf{Estimation:} (Cross-Validation) $\rightarrow$ \textit{How well will it perform?}
    \item \textbf{Uncertainty Quantification:} (Bootstrap) $\rightarrow$ \textit{How confident are we in that estimate?}
    \item \textbf{Theoretical Guarantee:} (VC Theory) $\rightarrow$ \textit{Why should we believe this is possible?}
    \item \textbf{Trust:} (Calibration) $\rightarrow$ \textit{Can we believe its own uncertainty?}
\end{enumerate}

Generalization is not a software bug to be hacked away; it is a statistical phenomenon to be understood, measured, and controlled. The frameworks for doing so were invented by statisticians. To build AI that is not just powerful but also \textbf{reliable}, \textbf{interpretable}, and \textbf{trustworthy}, one must first bow to the statistical principles that govern its behavior. \textbf{Every layer of intelligence—learning, reasoning, generalizing—rests upon this bedrock of statistical principles.}

%\section{Pillar V: Representation Learning}
% sections/pillar5_representation.tex
\section{Representation Learning and the Pursuit of Latent Structure (Pillar V)}
\label{sec:representation}

The curse of dimensionality is the fundamental obstacle to learning from complex data. The quintessential statistical solution is not to fit more complex models in the high-dimensional space, but to find a more meaningful, lower-dimensional representation of the data itself. This pillar demonstrates that the entire modern endeavor of representation learning—from word embeddings to vision transformers—is the direct descendant and scaling of a century of statistical research into latent variable models and dimensionality reduction. \textbf{There is no understanding of complex data without statistical representation learning}.

\begin{principle}[The Representation Principle]
Intelligent analysis requires transforming raw data into a representation that exposes the underlying latent factors of variation. This process of discovering simplifying representations is a core statistical problem.
\end{principle}

\subsection{The Statistical Ancestry: Linear Dimensionality Reduction}
The foundational statistical insight was that correlation structure in data implies a lower-dimensional representation.

\subsubsection{Pearson's Principal Component Analysis (PCA)}
In one of the most influential papers in all of statistics, \textcite{pearson1901} introduced PCA to address "the lines and planes of closest fit to systems of points in space." The goal was to find the orthogonal directions of maximum variance in the data, providing an optimal linear projection for dimensionality reduction.

The modern formulation via the eigendecomposition of the covariance matrix, $\mathbf{S} = \mathbf{V}\mathbf{\Lambda}\mathbf{V}^\top$, where the columns of $\mathbf{V}$ are the principal components, is the bedrock of multivariate statistics \parencite{jolliffe2002}. PCA is the statistical solution to finding the best low-rank approximation of the data matrix.

\subsubsection{Hotelling's Transformation and Factor Analysis}
Shortly after, \textcite{hotelling1933} provided the modern computational method for PCA. This work was parallel to the development of \textbf{Factor Analysis} \parencite{spearman1904, thurstone1947}, which posits that observed variables are linear combinations of some unobserved, latent factors plus noise:
\[
\mathbf{x} = \mathbf{W}\mathbf{z} + \boldsymbol{\epsilon}
\]
This model seeks not just an orthogonal projection, but a representation $\mathbf{z}$ that explains the covariance structure of $\mathbf{x}$. The philosophical shift from observed variables to latent constructs is one of statistics' greatest contributions to data science.

\subsection{The Nonlinear Evolution: Manifold Learning}
A profound limitation of PCA and Factor Analysis is their linearity. Much real-world data lies on or near a low-dimensional \textit{nonlinear} manifold embedded in a high-dimensional space. Statistics again provided the key insights.

\subsubsection{Multidimensional Scaling (MDS)}
MDS \parencite{torgerson1958} is a family of techniques that start with a matrix of pairwise \textit{dissimilarities} between data points and aim to find a representation in a lower-dimensional space that preserves these distances as well as possible. This emphasis on preserving geometric relationships, rather than just variance, was a critical conceptual leap.

\subsubsection{The Statistical Mechanics of Manifolds}
Modern nonlinear dimensionality reduction techniques are direct statistical descendants of MDS and kernel methods:
\begin{itemize}
    \item \textbf{Isomap} \parencite{tenenbaum2000} estimates the geodesic distance (the distance along the manifold) between points and uses MDS to find a representation that preserves these manifold distances.
    \item \textbf{Local Linear Embedding (LLE)} \parencite{roweis2000} assumes the manifold is locally linear and finds a representation that best preserves the local geometry of the data.
    \item \textbf{t-SNE} \parencite{maaten2008} uses a statistical measure of similarity (a Student-t kernel) to convert high-dimensional distances into probabilities and then minimizes the Kullback-Leibler divergence between the similarity distributions in the high- and low-dimensional spaces. This is pure statistical inference for the purpose of visualization.
\end{itemize}

\subsection{The Neural Embodiment: Autoencoders and Embeddings}
The deep learning revolution provided the function approximation power to learn highly nonlinear representations, but the goals remained thoroughly statistical.

\subsubsection{Autoencoders: Nonlinear PCA}
An autoencoder \parencite{rumelhart1985} is a neural network trained to copy its input to its output. The network is constrained by a "bottleneck" layer in the middle, forcing it to learn a compressed representation. A linear autoencoder with squared error loss learns the same subspace as PCA. A deep, nonlinear autoencoder learns a nonlinear manifold—it is the neural generalization of the statistical principle of dimensionality reduction.

\subsubsection{Word Embeddings: Distributional Semantics as Density Estimation}
The breakthrough of word2vec \parencite{mikolov2013} and GloVe \parencite{pennington2014} demonstrated that meaningful representations (embeddings) could be learned from data. The core insight is \textbf{distributional semantics}: "You shall know a word by the company it keeps" \parenquote{firth1957}. This is a statistical idea.
The algorithms work by modeling the probability of a word given its context (or vice versa), effectively performing a form of statistical density estimation in the space of word co-occurrences. The resulting embedding space captures semantic and syntactic relationships with stunning geometric regularity.

\subsubsection{Transformers as Representation Learning Machines}
The Transformer architecture \parencite{vaswani2017} can be viewed as the ultimate representation learning engine. Its self-attention mechanism allows every element in a sequence to integrate information from every other element, dynamically crafting a context-aware representation. The "output embeddings" of a pre-trained transformer like BERT \parencite{devlin2018} are powerful, task-agnostic representations of input data, precisely because the model has been trained on the statistical task of masked language modeling—a form of density estimation.

\subsection{Conclusion of Pillar V}
The quest for good representations is the quest for the true, simpler structure underlying complex data. This is a statistical quest, and its history is one of increasing nonlinearity and flexibility:
\begin{enumerate}
    \item \textbf{Linear Structure:} (PCA, Factor Analysis) $\rightarrow$ Find orthogonal directions of variance or latent factors.
    \item \textbf{Nonlinear Structure:} (MDS, Isomap, LLE, t-SNE) $\rightarrow$ Preserve distances or local geometry on a manifold.
    \item \textbf{Learnable Representations:} (Autoencoders, Embeddings, Transformers) $\rightarrow$ Use powerful function approximators to learn the mapping to a latent space directly from a statistical objective.
\end{enumerate}

From Pearson's planes of closest fit to the contextual embeddings of BERT, the goal has remained constant: to find a representation where the essential structure of the data is exposed and the noise is discarded. This is not a subfield of AI; it is the statistical heart of AI. \textbf{The entire architecture of artificial intelligence is built upon this statistical foundation} of discovering latent representations.

%\section{Pillar VI: Interpretability}
% sections/pillar6_interpretability_stat_ai.tex
\section{Trees, Ensembles, and the Statistical Pursuit of Interpretability (Pillar VI)}
\label{sec:interpretability}

A model's value is diminished if its decisions are opaque. The burgeoning field of Explainable AI (XAI) is, at its core, a applied statistical problem: how to quantify and communicate the contribution of input variables to a model's output. This pillar demonstrates that the statistical foundations of interpretability—from the recursive partitioning of trees to the model-agnostic techniques of modern XAI—were laid down by statisticians decades before the term "XAI" was coined. \textbf{There is no trustworthy AI without statistical interpretability}.

\begin{principle}[The Interpretability Principle]
Understanding a model's behavior requires decomposing its predictions into attributable components. The statistical frameworks for this decomposition provide the blueprint for explainable AI.
\end{principle}

\begin{principle}[The Interpretability Principle]
Understanding a model's behavior requires decomposing its predictions into attributable components. The statistical frameworks for this decomposition provide the blueprint for explainable AI.
\end{principle}

\subsection{The Decision Tree: A Statistical Model of Hierarchical Logic}

The decision tree is the most intuitively interpretable model, and its modern construction is a masterpiece of statistical reasoning.

\subsubsection{CART: A Statistical Framework for Building Trees}

The Classification and Regression Trees (CART) algorithm \parencite{breiman1984} provides a complete statistical methodology for building trees:
\begin{itemize}
    \item \textbf{Splitting Criterion:} The algorithm uses \textbf{impurity measures} to choose the best split at each node.
    \begin{itemize}
        \item \textbf{Gini Impurity:} $I_G(p) = 1 - \sum_{i=1}^C p_i^2$. A measure of statistical variance for categorical outcomes.
        \item \textbf{Entropy:} $I_E(p) = -\sum_{i=1}^C p_i \log p_i$. An information-theoretic measure of uncertainty.
        \item \textbf{Variance Reduction:} Used for regression trees, directly minimizing the statistical variance of the target within child nodes.
    \end{itemize}
    This transforms the problem of building a tree into a statistical optimization: find the split that maximizes the homogeneity (or minimizes the impurity) of the resulting subsets.
    \item \textbf{Pruning:} To avoid overfitting, CART uses \textbf{cost-complexity pruning}, which penalizes the tree's objective function for its number of terminal nodes (its complexity). This is a direct application of the statistical bias-variance tradeoff, formalizing Occam's razor for model structure.
\end{itemize}

\subsubsection{From Single Trees to Statistical Ensembles}

A single tree is interpretable but often high-variance. The statistical solution to high variance is aggregation.

\subsubsection{Bagging: Bootstrap Aggregation}

As introduced in Pillar IV, Bagging \parencite{breiman1996} is the application of the bootstrap to model building. By training many trees on bootstrap samples of the data and averaging their predictions, Bagging reduces variance and improves generalization. This is a purely statistical idea applied to create a stronger model.

\subsubsection{Random Forests: The Brilliant Statistical Intervention}

The Random Forest algorithm \parencite{breiman2001} adds a second layer of statistical wisdom to Bagging: \textbf{feature bagging}. When splitting a node, it only considers a random subset of features ($m \approx \sqrt{p}$). This de-correlates the trees in the ensemble further, dramatically reducing variance without a substantial increase in bias. This simple, statistical intervention is what makes Random Forests remarkably robust and accurate.

\subsubsection{Gradient Boosting: The Statistical Minimization of Residuals}

While bagging builds trees in parallel, boosting builds them sequentially. Algorithms like AdaBoost \parencite{freund1997} and Gradient Boosting Machines (GBM) \parencite{friedman2001} work by iteratively focusing on the examples that the current ensemble finds hardest to predict (the residuals). GBM explicitly frames this as a gradient descent optimization in function space, a statistical generalization of optimization. XGBoost \parencite{chen2016} and LightGBM are modern engineering implementations of this core statistical concept.

\subsection{Model Interpretation: Statistical Decomposition of Predictions}

Once a model is built, the next statistical problem is to explain its individual predictions.

\subsubsection{Feature Importance: A Statistical Measure of Contribution}

Tree-based models provide natural, statistical measures of feature importance:
\begin{itemize}
    \item \textbf{Gini Importance:} The total amount by which a feature reduces Gini impurity (or variance) across all splits in which it is used. This is an additive measure of a feature's total contribution to homogenizing the output.
    \item \textbf{Permutation Importance:} A model-agnostic method where the predictive power of a model is measured after randomly shuffling a single feature. The drop in performance indicates the feature's importance. This is a direct application of the statistical concept of randomization tests \parencite{fisher1935}.
\end{itemize}

\subsubsection{Partial Dependence Plots (PDPs)}

Introduced by \textcite{friedman2001}, PDPs show the marginal effect of one or two features on the predicted outcome of a model, averaging out the effects of all other features. This is literally visualizing the partial derivative of the model's prediction function, a concept from calculus, applied statistically by averaging over the data distribution.

\subsubsection{SHAP Values: Unifying Game Theory and Statistics}

The SHAP (SHapley Additive exPlanations) framework \parencite{lundberg2017} is the pinnacle of statistical interpretability. It roots model explanation in coalitional game theory \parencite{shapley1953}, where the prediction is the payout and the features are the players. The SHAP value for a feature is its average \textbf{marginal contribution} across all possible sequences (permutations) in which it could have been added to the model.

This is not just an analogy; it is a rigorous mathematical framework that satisfies important fairness properties. The calculation of SHAP values for tree ensembles \parencite{lundberg2020} is a brilliant piece of algorithmic statistics, making exact computation efficient.

\subsection{Conclusion of Pillar VI}

The path from a single, interpretable tree to a powerful, explainable ensemble is a journey through statistical innovation:
\begin{enumerate}
    \item \textbf{Interpretable Base Model:} (CART) $\rightarrow$ Uses statistical impurity measures to build a model with a clear decision structure.
    \item \textbf{Ensemble for Performance:} (Bagging, Random Forests, Boosting) $\rightarrow$ Uses statistical resampling and optimization to combine many weak models into one strong, robust one.
    \item \textbf{Explanation:} (Feature Importance, PDPs, SHAP) $\rightarrow$ Uses statistical measures of marginal contribution and averaging to decompose the ensemble's predictions into understandable components.
\end{enumerate}

The entire workflow is a continuous thread of statistical reasoning. The tools of XAI are not software hacks; they are the application of deep statistical concepts—variance, impurity, marginal effect, and fair allocation—to the problem of explaining complex models. \textbf{To build AI that is not just powerful but also reliable, interpretable, and trustworthy, one must first bow to the statistical principles that govern its behavior.}

%\section{Pillar VII: Causal Inference}
% sections/pillar7_causality_stat_ai.tex
\section{Causal Inference and the Leap from Prediction to Understanding (Pillar VII)}
\label{sec:causality}

The pinnacle of intelligence is not prediction, but understanding. While machine learning excels at finding patterns and correlations, it is famously said that "correlation does not imply causation." This pillar demonstrates that the rigorous mathematical framework for making that leap—for inferring causal relationships from data—is a monumental contribution of statistics. Modern causal machine learning does not replace this framework; it operationalizes it at scale. \textbf{There is no true intelligence without statistical causal inference}.

\begin{principle}[The Causal Principle]
Answering "what if" questions (counterfactuals) requires a formal model of interventions. Statistics provides this model and the methodologies to estimate its quantities from data.
\end{principle}

\begin{principle}[The Causal Principle]
Answering "what if" questions (counterfactuals) requires a formal model of interventions. Statistics provides this model and the methodologies to estimate its quantities from data.
\end{principle}

\subsection{The Structural Causal Model: A Framework for Reality}

The foundational breakthrough was to create a mathematical language for causality separate from association.

\subsubsection{From Graphs to Causality}

A Structural Causal Model (SCM) \parencite{pearl2009} consists of:
\begin{enumerate}
    \item \textbf{A Set of Variables} $\{X_1, ..., X_p\}$.
    \item \textbf{A Set of Functions} assigning each variable a value based on its parents and an exogenous noise variable: $X_i := f_i(\text{PA}(X_i), U_i)$.
    \item \textbf{A Directed Acyclic Graph (DAG)} representing the causal relationships, where an edge $X_j \rightarrow X_i$ means $X_j$ is a direct cause of $X_i$.
\end{enumerate}
This simple yet powerful formalism, developed by a computer scientist building on statistical concepts, provides the language to precisely define causal concepts.

\subsubsection{The Ladder of Causation}

Pearl's hierarchy \parencite{pearl2019} categorizes intelligent reasoning into three levels:
\begin{itemize}
    \item \textbf{Association} (Seeing): "What is?" e.g., \textit{What is the probability of disease given a symptom?} This is the domain of standard ML and statistics.
    \item \textbf{Intervention} (Doing): "What if?" e.g., \textit{What would the probability of disease be if I took aspirin?} This requires the $do$-calculus.
    \item \textbf{Counterfactuals} (Imagining): "What if I had acted differently?" e.g., \textit{Would I have recovered faster if I had taken the drug?} This is the highest level of causal reasoning.
\end{itemize}
This hierarchy clearly shows that contemporary ML operates primarily on the first rung, while the tools to climb higher were forged in statistics.

\subsection{The Toolbox for Causal Inference}

Statistics provides an arsenal of methods for estimating causal effects from imperfect data.

\subsubsection{Randomized Controlled Trials (RCTs): The Gold Standard}

The RCT is the quintessential statistical invention for causal inference. By randomly assigning subjects to treatment and control groups, it ensures that the two groups are statistically equivalent in expectation, allowing any difference in outcomes to be attributed to the treatment. The Average Treatment Effect (ATE) is estimated simply as:
\[
\text{ATE} = E[Y | do(T=1)] - E[Y | do(T=0)] \approx \bar{Y}_{\text{treat}} - \bar{Y}_{\text{control}}
\]
This design eliminates confounding, the central obstacle to causal inference.

\subsubsection{The Statistical Arsenal for Observational Data}

Since RCTs are often impractical, statistics developed methods for inferring causality from observational data:
\begin{itemize}
    \item \textbf{Propensity Score Matching} \parencite{rosenbaum1983}: Balances confounders between treated and untreated groups by matching each treated unit with an untreated unit that has a similar probability (propensity) of receiving the treatment. This mimics randomization.
    \item \textbf{Instrumental Variables (IV)} \parencite{angrist1996}: Uses a variable (the instrument) that affects the treatment but does not affect the outcome except through the treatment. This provides a clever way to isolate exogenous variation in the treatment.
    \item \textbf{Difference-in-Differences (DiD)}: Compares the change in outcomes over time between a treatment group and a control group. This controls for unobserved, time-invariant confounders.
    \item \textbf{Regression Discontinuity Design (RDD)}: Exploits a sharp cutoff in a continuous variable that assigns treatment. Units just on either side of the cutoff are assumed to be comparable, allowing for causal inference.
\end{itemize}
These are all \textit{statistical} methods designed to create quasi-experimental conditions from observational data.

\subsection{Causal Machine Learning: Scaling Statistical Principles}

Modern causal ML does not invent new principles; it applies statistical learning to estimate the functions and quantities defined by statistical causal theory.

\subsubsection{Meta-Learners: The Statistical Toolkit Formalized}

These frameworks combine any ML model with causal estimation:
\begin{itemize}
    \item \textbf{The S-Learner} (Single Learner): Trains a single model $\hat{\mu}(X, T)$ to predict the outcome from features and treatment. The treatment effect for an individual is estimated as $\hat{\tau}(X) = \hat{\mu}(X, 1) - \hat{\mu}(X, 0)$.
    \item \textbf{The T-Learner} (Two Learner): Trains two separate models, $\hat{\mu}_1(X)$ on the treated group and $\hat{\mu}_0(X)$ on the control group. Then, $\hat{\tau}(X) = \hat{\mu}_1(X) - \hat{\mu}_0(X)$.
    \item \textbf{The X-Learner} \parencite{kunzel2019}: A more sophisticated extension that often performs better, especially when treatment and control groups are of very different sizes.
\end{itemize}
These are not new causal theories; they are computational strategies for estimating the statistical quantity $\tau(x) = E[Y(1) - Y(0) | X=x]$.

\subsubsection{Doubly Robust Estimation}

A key statistical advancement combines outcome modeling (e.g., regression) with propensity score modeling \parencite{bang2005}:
\[
\hat{\tau}_{\text{DR}} = \frac{1}{N} \sum_i \left[ \frac{T_i Y_i}{\hat{e}(X_i)} - \frac{T_i - \hat{e}(X_i)}{\hat{e}(X_i)} \hat{\mu}_1(X_i) \right] - \left[ \frac{(1-T_i) Y_i}{1-\hat{e}(X_i)} - \frac{T_i - \hat{e}(X_i)}{1-\hat{e}(X_i)} \hat{\mu}_0(X_i) \right]
\]
This estimator is "doubly robust" because it remains consistent if \textit{either} the propensity score model $\hat{e}(x)$ \textit{or} the outcome models $\hat{\mu}_1(x), \hat{\mu}_0(x)$ are correctly specified. This is a statistical property guaranteeing robustness to model misspecification.

\subsection{Conclusion of Pillar VII}

The journey from prediction to causation is a journey through statistical innovation:
\begin{enumerate}
    \item \textbf{Formal Framework:} (SCMs, $do$-calculus) $\rightarrow$ Provides the language to define causal concepts mathematically.
    \item \textbf{Identification Strategies:} (RCTs, Matching, IV, DiD, RDD) $\rightarrow$ Provides the \textit{statistical} methods to isolate causal effects from data.
    \item \textbf{Estimation:} (Meta-Learners, Doubly Robust Estimation) $\rightarrow$ Uses machine learning to powerfully estimate the causal quantities defined by the framework.
\end{enumerate}

Causal machine learning is not a new field separate from statistics; it is the application of statistical learning to the problems formalized by statistical causal theory. The $do$-operator is the statistical bridge from the land of correlation to the promised land of causation. \textbf{Ignoring these roots risks building a fragile future; embracing them is the path to truly intelligent machines.}

%\section{Pillar VIII: Optimization}
% sections/pillar8_optimization_stat_ai.tex
\section{Optimization as Inference: The Blurring of Algorithm and Statistic (Pillar VIII)}
\label{sec:optimization}

The separation between the algorithm that finds a model (optimization) and the model itself (statistics) is an artificial one. This pillar reveals a profound unification: that many optimization procedures used in machine learning are, in fact, exact or approximate methods for statistical inference. The process of training a model is not just a computational hack; it is a statistical ritual for updating beliefs in the face of data. \textbf{There is no learning without statistical optimization}.

\begin{principle}[The Optimization-Inference Principle]
Many machine learning training algorithms can be reinterpreted as algorithms for performing statistical inference, blurring the line between optimization and probability.
\end{principle}

\begin{principle}[The Optimization-Inference Principle]
Many machine learning training algorithms can be reinterpreted as algorithms for performing statistical inference, blurring the line between optimization and probability.
\end{principle}

\subsection{Maximum Likelihood: The Original Optimization as Inference}
The most direct connection is the equivalence between maximizing likelihood and minimizing a specific loss.

\subsubsection{The Principle of Maximum Likelihood}

The principle of Maximum Likelihood Estimation (MLE), formalized by \textcite{fisher1922}, states that the parameters $\boldsymbol{\theta}$ that maximize the probability of the observed data are the "best" ones:
\[
\hat{\boldsymbol{\theta}}_{\text{MLE}} = \arg\max_{\boldsymbol{\theta}} \, p(\mathscr{D}_n \mid \boldsymbol{\theta})
\]
This is an optimization problem. For i.i.d. data, it is equivalent to minimizing the negative log-likelihood:
\[
\hat{\boldsymbol{\theta}}_{\text{MLE}} = \arg\min_{\boldsymbol{\theta}} \, \left( -\sum_{i=1}^n \log p(\mathbf{x}_i \mid \boldsymbol{\theta}) \right)
\]
This simple transformation frames probability maximization as loss minimization, the standard paradigm of ML.

\subsubsection{The Loss-Likelihood Dictionary}

This equivalence creates a direct mapping between common loss functions and statistical models:
\begin{itemize}
    \item \textbf{Mean Squared Error} $\leftrightarrow$ MLE under a Gaussian noise model.
    \item \textbf{Cross-Entropy Loss} $\leftrightarrow$ MLE for a categorical distribution (logistic regression).
    \item \textbf{Mean Absolute Error} $\leftrightarrow$ MLE under a Laplacian noise model.
\end{itemize}
Choosing a loss function is implicitly choosing a probabilistic model for the data.

\subsection{The Bayesian Reformation: Inference over Point Estimates}

The Bayesian approach rejects the idea of a single "best" parameter, instead maintaining a full distribution over possibilities.

\subsubsection{From Optimization to Integration}

Bayesian inference flips the problem from optimization to integration. We seek the posterior distribution:
\[
p(\boldsymbol{\theta} \mid \mathscr{D}_n) = \frac{p(\mathscr{D}_n \mid \boldsymbol{\theta}) p(\boldsymbol{\theta})}{p(\mathscr{D}_n)}
\]
The denominator, $p(\mathscr{D}_n) = \int p(\mathscr{D}_n \mid \boldsymbol{\theta}) p(\boldsymbol{\theta}) d\boldsymbol{\theta}$, is the computationally challenging integral. This is not an optimization problem but a problem of \textbf{approximate integration}.

\subsubsection{Variational Inference: Optimization for Approximation}

Variational Inference (VI) \parencite{jordan1999} cleverly turns the integration problem back into an optimization one. It introduces a family of simpler distributions $q(\boldsymbol{\theta} \mid \boldsymbol{\phi})$ and finds the member that is closest to the true posterior:
\[
q^*(\boldsymbol{\theta}) = \arg\min_{q \in \mathcal{Q}} \text{KL}\left( q(\boldsymbol{\theta}) \parallel p(\boldsymbol{\theta} \mid \mathscr{D}_n) \right)
\]
Minimizing the Kullback-Leibler divergence is an optimization problem whose solution provides an approximate posterior. This makes Bayesian inference tractable for complex models.

\subsubsection{Expectation-Maximization: Optimization via Latent Variables}

The EM algorithm \parencite{dempster1977} is a quintessential example of optimization as inference for models with latent variables $\mathbf{z}$. It alternates between:
\begin{enumerate}
    \item \textbf{E-step:} Compute the expected value of the log-likelihood, $Q(\boldsymbol{\theta} \mid \boldsymbol{\theta}^{(t)}) = E_{\mathbf{z} \mid \mathbf{x}, \boldsymbol{\theta}^{(t)}}[\log p(\mathbf{x}, \mathbf{z} \mid \boldsymbol{\theta})]$. This is an inference step.
    \item \textbf{M-step:} Find the parameters that maximize this expectation: $\boldsymbol{\theta}^{(t+1)} = \arg\max_{\boldsymbol{\theta}} Q(\boldsymbol{\theta} \mid \boldsymbol{\theta}^{(t)})$. This is an optimization step.
\end{enumerate}
EM is a procedure that iteratively performs inference to enable optimization, which in turn improves the next inference step.

\subsection{Stochastic Gradient Descent: The Statistical Workhorse}

The algorithm that powers modern deep learning is itself deeply statistical.

\subsubsection{Robbins-Monro: The Statistical Foundation of SGD}

The Stochastic Gradient Descent (SGD) algorithm is a direct application of the \textbf{Robbins-Monro algorithm} \parencite{robbins1951} for stochastic approximation. Robbins and Monro provided the theoretical conditions under which an iterative update of the form:
\[
\boldsymbol{\theta}^{(t+1)} = \boldsymbol{\theta}^{(t)} - \eta_t \nabla_{\boldsymbol{\theta}} L(\boldsymbol{\theta}^{(t)}; \mathbf{x}_i, y_i)
\]
will converge to the optimum $\boldsymbol{\theta}^*$. Their work guaranteed that using noisy, unbiased estimates of the gradient (from mini-batches) is a statistically valid way to solve the optimization problem.

\subsubsection{SGD as Approximate Bayesian Inference}

A remarkable line of research \parencite{mandt2017, smith2018} has shown that under certain conditions, SGD itself can be interpreted as an approximate Bayesian inference algorithm.
\begin{itemize}
    \item The trajectory of SGD parameters can be seen as sampling from a distribution.
    \item Adding the right amount of noise to the SGD updates (e.g., via a specific learning rate schedule) can make the algorithm converge to the true Bayesian posterior $p(\boldsymbol{\theta} \mid \mathscr{D}_n)$, not just a point estimate.
    \item The common practice of using a high learning rate early and decaying it resembles a annealing process used in Bayesian sampling.
\end{itemize}
This means that when we train a neural network with SGD, we may be performing implicit, approximate Bayesian inference, whether we realize it or not.

\subsection{Conclusion of Pillar VIII}

This pillar demonstrates that the boundary between optimization and inference is not just porous—it is often nonexistent:
\begin{enumerate}
    \item \textbf{MLE} directly frames statistical inference as an optimization problem.
    \item \textbf{Variational Inference} frames Bayesian inference as an optimization problem over distributions.
    \item \textbf{The EM Algorithm} seamlessly intertwines inference and optimization steps.
    \item \textbf{Stochastic Gradient Descent}, the engine of deep learning, is both a stochastic optimization algorithm and a form of approximate Bayesian inference.
\end{enumerate}

The algorithm and the statistic are two sides of the same coin. The process of learning is the process of inference, and the process of inference is very often an optimization. This is one of the deepest and most beautiful unifications that statistics provides to machine learning. \textbf{To build AI that is not just powerful but also reliable, interpretable, and trustworthy, one must first bow to the statistical principles that govern its behavior.}

%\section{Pillar IX: Unification}
% sections/pillar9_unification_stat_ai.tex
\section{Unification, Universality, and the Statistical Fabric of Learning (Pillar IX)}
\label{sec:unification}

The previous pillars have traced the statistical lineage of specific AI paradigms. This final pillar ascends to a higher vantage point, revealing the profound unifying principles that weave these paradigms into a coherent statistical fabric. We explore how core statistical concepts—like ensemble methods, kernel tricks, and universal function approximation—manifest across AI, creating a surprising cohesion between methods once thought distinct. \textbf{The statistical footprints permeate the entire landscape of machine learning and artificial intelligence} not as isolated paths, but as a unified network of ideas.

\begin{principle}[The Unification Principle]
Beneath the apparent diversity of machine learning algorithms lies a deep unity of statistical principles. Ensemble methods, kernel learning, and universal approximation are all expressions of fundamental statistical concepts.
\end{principle}

\begin{principle}[The Unification Principle]
Beneath the apparent diversity of machine learning algorithms lies a deep unity of statistical principles. Ensemble methods, kernel learning, and universal approximation are all expressions of fundamental statistical concepts.
\end{principle}

\subsection{The Statistical Ensemble: Wisdom of the Crowd}

The principle of combining models is a statistical idea of stunning power and generality, far beyond Random Forests.

\subsubsection{Bayesian Model Averaging (BMA): The Gold Standard}

BMA \parencite{hoeting1999} is the rigorous statistical framework for ensemble learning. Instead of choosing one model, it averages predictions from all possible models, weighted by their posterior probability:
\[
p(y^* | \mathbf{x}^*, \mathscr{D}_n) = \sum_{m \in \mathcal{M}} p(y^* | \mathbf{x}^*, m)  p(m | \mathscr{D}_n)
\]
This is the optimal way to account for model uncertainty. Practical ML ensembles (bagging, boosting, stacking) are often approximations or efficient implementations of this Bayesian ideal.

\subsubsection{Stacking: Learning the Combiner}

Stacking (Stacked Generalization) \parencite{wolpert1992} takes ensemble learning a step further by using a meta-model to learn how to best combine the predictions of base learners. This is a form of \textbf{learned model averaging}, where the combination weights are not just based on simple performance but on a learned function of the input data itself. This is a statistical concept—optimal combination—scaled via machine learning.

\subsubsection{The Deep Learning Ensemble}

Even in deep learning, ensembles are a powerful tool. Techniques like \textbf{Monte Carlo Dropout} \parencite{gal2016} effectively create an ensemble of networks during inference by performing multiple forward passes with dropout enabled. This provides uncertainty estimates and is a clever, practical implementation of a Bayesian ensemble.

\subsection{The Kernel: The Statistical Trick of the Century}

The "kernel trick" is a breathtaking statistical idea that elegantly solves the curse of dimensionality.

\subsubsection{Beyond Gaussian Processes: The Kernel Zoo}

While Gaussian Processes are a canonical kernel method, the statistical theory of Reproducing Kernel Hilbert Spaces (RKHS) \parencite{aronszajn1950} underpins a vast array of algorithms:
\begin{itemize}
    \item \textbf{Support Vector Machines (SVMs)} \parencite{cortes1995}: For classification and regression, which find the optimal separating hyperplane in a high-dimensional feature space defined implicitly by a kernel.
    \item \textbf{Kernel PCA} \parencite{scholkopf1997}: A nonlinear extension of PCA that performs dimensionality reduction in a feature space.
    \item \textbf{Kernel Ridge Regression}: Applies ridge regression in the RKHS, providing a powerful nonlinear regression technique.
\end{itemize}
The kernel $k(\mathbf{x}, \mathbf{x}') = \langle \phi(\mathbf{x}), \phi(\mathbf{x}') \rangle$ is a measure of similarity, and kernel methods can be seen as sophisticated \textbf{nonparametric smoothing} techniques that operate by comparing data points to each other.

\subsubsection{Neural Tangents and Gaussian Universality}

A groundbreaking modern unification has emerged: \textbf{Neural Tangent Kernel (NTK)} theory \parencite{jacot2018}. It reveals that in the limit of infinite width, the behavior of a deep neural network during training by gradient descent is equivalent to kernel regression with a specific kernel (the NTK). This stunning result bridges the seemingly disparate worlds of:
\begin{itemize}
    \item \textbf{Neural Networks}: Represented by their function approximation and training dynamics.
    \item \textbf{Kernel Methods}: Represented by the NTK and the statistical theory of RKHS.
    \item \textbf{Gaussian Processes}: As the network width approaches infinity, the prior over functions defined by randomly initialized weights becomes a Gaussian Process.
\end{itemize}
This "Gaussian universality" demonstrates that deep learning, in its modern idealized form, is fundamentally a statistical kernel method. The \textbf{statistical fingerprints} are not just on the surface; they form the very theoretical bedrock of why deep learning works.

\subsection{Predictive Analytics: The Statistical Engine of Industry}

The vast field of predictive analytics is the applied embodiment of statistical learning, focused on operationalizing prediction.

\subsubsection{The CRISP-DM Framework}

The Cross-Industry Standard Process for Data Mining (CRISP-DM) \parencite{chapman1999} is the dominant workflow for predictive modeling. Its phases—Business Understanding, Data Understanding, Data Preparation, Modeling, Evaluation, Deployment—are a formalization of the statistical data analysis cycle. The entire process is a pipeline for managing statistical uncertainty and validating models against business objectives.

\subsubsection{MLOps: Statistical Monitoring at Scale}

The modern practice of MLOps (Machine Learning Operations) is applied statistics for the AI lifecycle. It involves:
\begin{itemize}
    \item \textbf{Data Drift Detection:} Using statistical tests (e.g., Kolmogorov-Smirnov, $\chi^2$) to monitor if the distribution of incoming data $p_{\text{new}}(\mathbf{x})$ diverges from the training data $p_{\text{train}}(\mathbf{x})$.
    \item \textbf{Model Performance Monitoring:} Tracking metrics over time and setting up statistical alerting for significant drops in performance.
    \item \textbf{Canary Deployments and A/B Testing:} Using statistical hypothesis testing to safely roll out new model versions.
\end{itemize}
MLOps is the engineering implementation of the statistical principle that a model is only valid under the distribution it was trained on, and requires continuous statistical validation.

\subsection{The Symbiosis: Statistics as the Brain, Computation as the Brawn} % <-- NEW CRITICAL SUBSECTION

A complete narrative must acknowledge the indispensable role of computer engineering. The statistical principles we have extolled would remain elegant but largely impotent theories without the computational power to realize them at scale. The rise of specialized hardware, epitomized by NVIDIA's GPUs, is not a refutation of our thesis but its ultimate validation. It provides the brawn to the statistical brain.

This is a profound symbiosis, echoing the old adage: \textit{"Theory without practice is lame. Practice without theory is blind."}
\begin{itemize}
    \item \textbf{Statistics is the "Theory":} It provides the objective functions (MLE), the inference frameworks (Bayesian updating), the validation methods (cross-validation), and the theoretical guarantees (VC theory). It is the \textit{intelligence}—the "why" and the "what".
    \item \textbf{Computer Science is the "Practice":} It provides the scalable algorithms (SGD, backpropagation), the hardware architectures (GPUs, TPUs), and the software ecosystems (TensorFlow, PyTorch). It is the \textit{engine}—the "how" and the "what with".
\end{itemize}

The dominance of hardware pioneers like NVIDIA does not contradict our thesis; it completes it. They built the powerful engines, but statistics provides the navigation system and the maps. The statistical principles we have outlined would remain elegant but impotent abstractions without the
computational power to realize them. Conversely, raw computational power, without statistical guidance, is brute force without direction—a danger the field has encountered in the pursuit of scale without understanding. The future belongs to those who master both

The modern AI revolution is the story of this marriage: statistical thought, conceived decades or even centuries ago, meeting computational power that is finally capable of breathing life into it. The GPU does not replace the statistician; it \textit{empowers} the statistician's ideas to solve problems previously thought intractable.

\subsection{Conclusion: The Unified Statistical Fabric of AI}

This final pillar reveals that the eight previous pillars are not separate columns but intertwined threads in a single tapestry:
\begin{itemize}
    \item \textbf{Ensemble Methods} (This Pillar) unite the inference of Pillar I with the resampling of Pillar IV.
    \item \textbf{Kernel Methods} unite the representation learning of Pillar V with the function approximation of deep learning.
    \item \textbf{The NTK} provides a deep theoretical unification of neural networks and statistical kernel theory.
    \item \textbf{Predictive Analytics} is the applied manifestation of the entire statistical learning cycle, from data understanding (Pillar IV) to deployment.
    \item \textbf{The Symbiosis} unites the statistical "brain" with the computational "brawn," creating the modern AI ecosystem.
\end{itemize}

The journey from a single algorithm to the entire ecosystem of AI is a continuous spectrum of statistical reasoning. The field is not a collection of isolated tricks but a coherent edifice built on a foundation of statistical principles. This is the powerful, unifying message of our manifesto: \textbf{From its theoretical roots to its modern algorithms, artificial intelligence is applied statistics at scale}.

% sections/empirical_imperative.tex
\section{The Empirical Imperative: From Theoretical Lineage to Trustworthy Systems}
\label{sec:empirical_imperative}

The nine pillars we have erected present a compelling historical and theoretical argument for the statistical soul of AI. However, the ultimate validation of any scientific framework is its utility in practice. A theory that cannot explain or improve real-world outcomes remains a philosophical exercise. In this section, we bridge the gap between lineage and liability, theory and trust. We present a brief empirical manifesto demonstrating that ignoring the statistical foundation leads to fragile, unreliable systems, while embracing it paves the path to robust, trustworthy intelligence.

\subsection{The Peril of Statistical Neglect: Brittle Predictors and Silent Failures}

The modern AI landscape is littered with examples of models that excel on benchmark datasets but fail catastrophically in the wild. This brittleness is not a random bug; it is a direct symptom of a statistically-naive approach focused solely on point prediction.

Consider the phenomenon of \textbf{adversarial attacks} \cite{Goodfellow2014ExplainingAH, Szegedy2013Intriguing}. A deep neural network, trained to state-of-the-art accuracy on a dataset like CIFAR-10 or ImageNet via maximum likelihood estimation (Pillar I), can have its predictions completely reversed by an imperceptibly small, carefully crafted perturbation to an input image. This model, a master of interpolation on the training manifold, possesses no statistical understanding of its own epistemic limits. It has learned complex correlations but cannot quantify its uncertainty. When presented with an out-of-distribution input—even a maliciously constructed one—it provides a high-confidence, catastrophicly wrong answer. It fails, and it fails \textit{silently}.

This problem extends beyond adversarial noise to natural distribution shifts. A model deployed in a clinical setting may perform well on data from one hospital but fail on data from another due to differences in equipment or patient demographics. A model evaluating the statistical properties of the input data, perhaps through the lens of the density estimation principles in Pillar II (e.g., using a simple generative model to flag low-likelihood inputs), could alert users to this discrepancy. Without this, the model blindly applies its learned function, unaware that it is operating far from its training domain.

\subsection{The Power of Statistical Embrace: Robustness and Self-Awareness}

In stark contrast, systems built with statistical principles at their core demonstrate a capacity for robustness and self-awareness. Let us revisit the adversarial example, but now with a \textbf{Bayesian Neural Network (BNN)} \cite{mackay1992, neal1993} or a model utilizing \textbf{Mote Carlo Dropout} \cite{gal2016} as a practical Bayesian approximation.

Instead of producing a single point estimate, these models produce a \textit{predictive distribution}. When presented with the same adversarial example that fooled the deterministic network, the BNN's predictive entropy will be significantly higher, and its maximum predictive probability will be much lower. It effectively flags its own confusion, signaling low confidence. This is statistical uncertainty quantification (Pillar I) in action, providing a built-in alarm system for anomalous or malicious inputs. The model fails \textit{gracefully}, informing its users of its limitations and preventing a silent, high-consequence error.

This statistical embrace is equally critical for model assessment and deployment (Pillar IV). A performance metric calculated on a single, static test set provides a fragile, point-estimate guarantee. In contrast, applying the \textbf{bootstrap} \cite{efron1979, efron1994}—a quintessential statistical resampling technique—to model performance allows us to generate confidence intervals for accuracy, precision, or any other metric. This provides a far richer, more trustworthy assessment of how the model will perform on the true, unseen population. It moves the evaluation from "the model is 95\% accurate" to "we are 95\% confident the model's accuracy lies between 92\% and 97\%," a fundamentally more informative and scientifically rigorous statement.

Furthermore, the entire practice of \textbf{MLOps} relies on statistical hypothesis testing to monitor for data and concept drift, ensuring that the foundational assumption of identical training and deployment distributions holds. Without these statistical guardrails, model performance can decay silently, leading to significant financial or operational damage.

\subsection{Conclusion: The Indispensable Footprint}

The statistical footprints we have traced are not merely historical artifacts or academic curiosities. They are the very features that separate a clever pattern-matching algorithm from a reliable intelligent system. The path forward for AI is not to abandon these principles in pursuit of ever-larger, more opaque models, but to integrate them more deeply.

\begin{itemize}
    \item \textbf{Uncertainty Quantification} (Pillar I) is not a luxury; it is a prerequisite for deployment in high-stakes environments like medicine and autonomous driving.
    \item \textbf{Density Estimation} (Pillar II) provides the machinery for novelty and anomaly detection, safeguarding systems against irrelevant or malicious data.
    \item \textbf{Resampling and Validation} (Pillar IV) transform brittle point estimates into robust, dependable measures of performance.
    \item \textbf{Causal Reasoning} (Pillar VII) is the only path from mere prediction to actionable understanding and intervention.
\end{itemize}

The empirical imperative is clear: to build AI that is not just powerful but also \textbf{reliable, interpretable, and trustworthy}, we must consciously and deliberately architect it upon the statistical foundation that gave it life. The choice is between building brittle idols of correlation or robust engines of understanding.

% ... More Pillars ...

%\section{Conclusion: Reclaiming the Narrative}
% sections/conclusion.tex
\section{Conclusion: Reclaiming the Narrative -- The Statistical Imperative for AI}
\label{sec:conclusion}

We have journeyed through nine pillars that form the foundation of artificial intelligence, and at the heart of each, we have found not a novel algorithmic invention, but a profound statistical principle. From the inferential frameworks that govern learning to the causal models that promise true understanding, from the optimization of parameters to the interpretation of predictions, the story is the same: \textbf{the theoretical and methodological core of AI is, and has always been, statistical}.

This is not a claim of mere historical precedence. It is an argument about essential identity. The algorithms have scaled, the compute has grown exponentially, and the applications have multiplied, but the underlying grammar of learning remains unchanged. The footprints of Pearson, Fisher, Tukey, Efron, Breiman, Pearl, and countless other statisticians are not faint impressions on the path; they are the very road upon which modern AI travels.

\subsection{Synthesis of the Argument}

Our exploration reveals a coherent statistical fabric woven through AI:
\begin{itemize}
    \item \textbf{Learning is Inference} (Pillars I \& VIII): The process of training a model, whether through maximizing likelihood or variational inference, is a statistical ritual of updating beliefs from data.
    \item \textbf{Understanding Requires Representation} (Pillars II \& V): Whether discovering clusters, estimating densities, or learning embeddings, the goal is to find a meaningful latent structure—a century-old statistical problem.
    \item \textbf{Value is Determined by Generalization} (Pillar IV): The entire framework for evaluating models—cross-validation, bootstrapping, theoretical bounds—is a statistical apparatus for estimating performance beyond the sample.
    \item \textbf{Trust is Built on Interpretation} (Pillar VI): Explaining a model's behavior through feature importance, partial dependence, and Shapley values is an exercise in statistical decomposition.
    \item \textbf{Intelligence Demands Causation} (Pillar VII): The move from predictive patterns to actionable interventions is enabled by the formal language of structural causal models and the statistical ingenuity of quasi-experimental methods.
    \item \textbf{Unification is Achieved Through Principles} (Pillar IX): Ensemble methods, kernel tricks, and optimization dynamics are all expressions of deeper statistical concepts like model averaging, similarity metrics, and stochastic approximation.
\end{itemize}

\subsection{The Symbiotic Imperative: Theory and Practice} % <-- NEW BALANCING SUBSECTION

In reclaiming the statistical soul of AI, we must also celebrate its computational body. The monumental achievements of modern AI are the product of a profound symbiosis, echoing the adage that \textit{"theory without practice is lame; practice without theory is blind."}

\begin{itemize}
    \item \textbf{Statistics is the Brain:} It provides the fundamental questions, the inferential frameworks, the uncertainty quantification, and the theoretical guarantees. It is the \textit{intelligence}—the "why" and the "what for".
    \item \textbf{Computer Science is the Brawn:} It provides the scalable algorithms, the hardware architectures (GPUs, TPUs), and the software ecosystems that transform statistical theory into working intelligence. It is the \textit{engine}—the "how" and the "what with".
\end{itemize}

The dominance of hardware pioneers like NVIDIA does not contradict our thesis; it completes it. They built the powerful engines, but statistics provides the navigation system and the maps. The statistical principles we have outlined would remain elegant but impotent abstractions without the computational power to realize them. Conversely, raw computational power, without statistical guidance, is brute force without direction—a danger the field has encountered in the pursuit of scale without understanding. The future belongs to those who master both.

\subsection{Implications and Call to Action} % <-- Now this follows naturally from the balanced view

Recognizing this dual foundation is not an academic exercise; it is an imperative with profound practical consequences:

\begin{enumerate}
    \item \textbf{For Education:} The curriculum for the next generation of AI practitioners must be deeply rooted in statistical theory \textit{and} computational practice. We must teach the statistical \textit{why} behind the algorithmic \textit{how}. A solid grasp of probability, inference, and experimental design is non-negotiable for building robust, fair, and effective AI systems.
    \item \textbf{For Research:} The field must move beyond its current focus on scale-alone benchmarks. Grounding research in statistical theory provides a compass for innovation. It encourages us to ask not just "does it work?" but "why does it work?" and "under what conditions will it fail?".
    \item \textbf{For Practice:} Developers and engineers must adopt the statistical mindset of uncertainty quantification. Deploying a model without confidence intervals, without tests for data drift, and without a measure of its reliability is not engineering; it is recklessness. Statistical thinking is the primary tool for managing the risk of AI deployment.
\end{enumerate}

\subsection{A Look Forward: The Statistical Future of AI}
The next frontiers of AI will be conquered not by abandoning this statistical foundation, but by embracing it more deeply in partnership with computational advances.
\begin{itemize}
    \item \textbf{Causal AI:} The integration of causal reasoning into deep learning represents the most promising path toward truly intelligent systems that can reason about interventions and counterfactuals.
    \item \textbf{Uncertainty-Aware AI:} The development of models that can honestly say "I don't know" will be crucial for high-stakes applications in medicine, autonomous systems, and policy.
    \item \textbf{Resource-Constrained AI:} Statistical design of experiments and active learning will be key to developing AI that can learn efficiently without requiring astronomical amounts of data and compute.
\end{itemize}

In closing, we return to our central thesis, now fully evidenced and contextualized: \textbf{There is no machine learning without statistical learning; no artificial intelligence without statistical thought}. The narrative that AI is a product of computer science alone is a historical oversimplification that obscures its true nature. It is time to reclaim the narrative. Artificial intelligence is applied statistics, powered by computation, at a revolutionary scale. Embracing this truth is not a diminishing of AI's achievements; it is the key to its future progress, its reliability, and its beneficial integration into human society.

% Print the bibliography
\printbibliography

% Optional: Appendices for the large figures
\begin{appendix}

\section{Statistical Lineage of AI: A Timeline}
% ====== CORRECTED TIMELINE ======

%%%%%%%
\begin{figure}[htbp]
\centering
\caption{Statistical Lineage of Artificial Intelligence}
\label{fig:timeline}
\begin{tikzpicture}[
    timeline/.style={thick, ->, black!60},
    event/.style={rectangle, align=center, font=\tiny, text width=1.8cm, minimum height=0.5cm},
    statEvent/.style={event, fill=orange!20, draw=orange!50, rounded corners=2pt},
    aiEvent/.style={event, fill=blue!20, draw=blue!50, rounded corners=2pt},
    year/.style={font=\tiny\itshape, above=0.1cm}
]

% Draw timeline
\draw[timeline] (0,0) -- (17,0);
\node[anchor=west, font=\tiny] at (17,0.3) {Time};

% Statistical events (below timeline)
\node[statEvent] (pearson) at (0.5,-1) {Pearson\\PCA};
\node[year] at (pearson.north) {1901};

\node[statEvent] (fisher) at (2,-1) {Fisher\\MLE};
\node[year] at (fisher.north) {1922};

\node[statEvent] (robbins) at (3.5,-1) {Robbins-Monro\\SGD};
\node[year] at (robbins.north) {1951};

\node[statEvent] (kalman) at (5,-1) {Kalman\\Filter};
\node[year] at (kalman.north) {1960};

\node[statEvent] (parzen) at (6.5,-1) {Parzen\\KDE};
\node[year] at (parzen.north) {1962};

\node[statEvent] (efron) at (8,-1) {Efron\\Bootstrap};
\node[year] at (efron.north) {1979};

\node[statEvent] (breiman) at (9.5,-1) {Breiman\\CART/RF};
\node[year] at (breiman.north) {1984/2001};

\node[statEvent] (pearl) at (11,-1) {Pearl\\SCMs};
\node[year] at (pearl.north) {2000};

\node[statEvent] (goodfellow) at (14,-1) {Goodfellow\\GANs};
\node[year] at (goodfellow.north) {2014};

\node[statEvent] (vaswani) at (16,-1) {Vaswani et al.\\Transformers};
\node[year] at (vaswani.north) {2017};

% AI events (above timeline)
\node[aiEvent] (autoenc) at (0.5,1) {Autoencoders};
\node[aiEvent] (loss) at (2,1) {DL Loss\\Functions};
\node[aiEvent] (sgd) at (3.5,1) {SGD\\Optimizer};
\node[aiEvent] (rnn) at (5,1) {RNNs};
\node[aiEvent] (anomaly) at (6.5,1) {Anomaly\\Detection};
\node[aiEvent] (bagging) at (8,1) {Bagging\\Ensembles};
\node[aiEvent] (trees) at (9.5,1) {Tree-based\\ML};
\node[aiEvent] (causal) at (11,1) {Causal\\ML};
\node[aiEvent] (genai) at (14,1) {Generative\\AI};
\node[aiEvent] (llm) at (16,1) {LLMs};

% Connect statistics to AI
\foreach \s/\a in {pearson/autoenc, fisher/loss, robbins/sgd, kalman/rnn, parzen/anomaly, efron/bagging, breiman/trees, pearl/causal, goodfellow/genai, vaswani/llm} {
    \draw[->, dashed, gray, thin] (\s) -- (\a);
}

% Labels
\node[font=\footnotesize\bfseries, orange] at (8.5,-2) {Statistical Foundations};
\node[font=\footnotesize\bfseries, blue] at (8.5,2) {AI Manifestations};

\end{tikzpicture}
\end{figure}
%%%%%

\section{Three part timeline}

% ====== PART 1: THE FOUNDATIONAL AGE (1901-1960) ======
\begin{figure}[htbp]
\centering
\begin{tikzpicture}[
    timeline/.style={thick, ->, black!60},
    event/.style={rectangle, align=center, font=\tiny, text width=1.6cm, minimum height=0.6cm},
    statEvent/.style={event, fill=orange!20, draw=orange!50, rounded corners=2pt},
    aiEvent/.style={event, fill=blue!20, draw=blue!50, rounded corners=2pt},
    pillarLabel/.style={font=\tiny\bfseries, rotate=90, anchor=center}
]

% Draw timeline
\draw[timeline] (0,0) -- (10.5,0);
\node[anchor=west, font=\tiny] at (10.5,0.3) {Time};

% Title
\node[above, font=\small\bfseries] at (5.25, 2.2) {The Foundational Age (1901-1960)};
\node[above, font=\footnotesize] at (5.25, 1.9) {The Birth of Core Statistical Ideas};

% Pillar Regions (more compact)
\draw[orange, fill=orange!3, very thin] (0.3,-1.5) rectangle (2.7,1.5); 
\node[pillarLabel] at (0.1, 0) {Pillar V};

\draw[blue, fill=blue!3, very thin] (2.9,-1.5) rectangle (5.1,1.5); 
\node[pillarLabel] at (2.7, 0) {Pillar I};

\draw[red!70, fill=red!3, very thin] (5.3,-1.5) rectangle (7.2,1.5); 
\node[pillarLabel] at (5.1, 0) {Pillar VIII};

\draw[green!60!black, fill=green!3, very thin] (7.4,-1.5) rectangle (9.7,1.5); 
\node[pillarLabel] at (7.2, 0) {Pillar III};

% Events - Statistics (below timeline)
\node[statEvent] (pearson) at (1.5,-1) {Pearson\\PCA (1901)};
\node[statEvent] (fisher) at (4,-1) {Fisher\\MLE (1922)};
\node[statEvent] (robbins) at (6.2,-1) {Robbins-Monro\\SGD (1951)};
\node[statEvent] (kalman) at (8.5,-1) {Kalman\\Filter (1960)};

% Events - AI (above timeline)
\node[aiEvent] (autoenc) at (1.5,1) {Autoencoders};
\node[aiEvent] (mleloss) at (4,1) {DL Loss\\Functions};
\node[aiEvent] (sgd) at (6.2,1) {SGD\\Optimizer};
\node[aiEvent] (rnn) at (8.5,1) {RNNs};

% Connections
\foreach \s/\a in {pearson/autoenc, fisher/mleloss, robbins/sgd, kalman/rnn} {
    \draw[->, dashed, gray, thin] (\s) -- (\a);
}

% Labels
\node[font=\tiny, orange] at (5.25,-1.8) {Statistical Foundations};
\node[font=\tiny, blue] at (5.25,1.8) {AI Manifestations};

\end{tikzpicture}
\end{figure}

%%%%%%%%%

% ====== PART 2: THE COMPUTATIONAL AGE (1960-2000) ======
\begin{figure}[htbp]
\centering
\begin{tikzpicture}[
    timeline/.style={thick, ->, black!60},
    event/.style={rectangle, align=center, font=\tiny, text width=1.6cm, minimum height=0.6cm},
    statEvent/.style={event, fill=orange!20, draw=orange!50, rounded corners=2pt},
    aiEvent/.style={event, fill=blue!20, draw=blue!50, rounded corners=2pt},
    pillarLabel/.style={font=\tiny\bfseries, rotate=90, anchor=center}
]

% Draw timeline
\draw[timeline] (0,0) -- (10.5,0);
\node[anchor=west, font=\tiny] at (10.5,0.3) {Time};

% Title
\node[above, font=\small\bfseries] at (5.25, 2.2) {The Computational Age (1960-2000)};
\node[above, font=\footnotesize] at (5.25, 1.9) {The Bridge to Algorithms and Practice};

% Pillar Regions (compact)
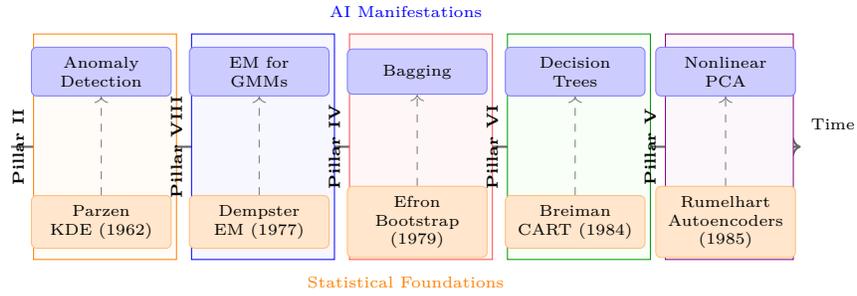
\draw[orange, fill=orange!3, very thin] (0.3,-1.5) rectangle (2.2,1.5); 
\node[pillarLabel] at (0.1, 0) {Pillar II};

\draw[blue, fill=blue!3, very thin] (2.4,-1.5) rectangle (4.3,1.5); 
\node[pillarLabel] at (2.2, 0) {Pillar VIII};

\draw[red!70, fill=red!3, very thin] (4.5,-1.5) rectangle (6.4,1.5); 
\node[pillarLabel] at (4.3, 0) {Pillar IV};

\draw[green!60!black, fill=green!3, very thin] (6.6,-1.5) rectangle (8.5,1.5); 
\node[pillarLabel] at (6.4, 0) {Pillar VI};

\draw[violet, fill=violet!3, very thin] (8.7,-1.5) rectangle (10.4,1.5); 
\node[pillarLabel] at (8.5, 0) {Pillar V};

% Events - Statistics (below timeline)
\node[statEvent] (parzen) at (1.2,-1) {Parzen\\KDE (1962)};
\node[statEvent] (demster) at (3.3,-1) {Dempster\\EM (1977)};
\node[statEvent] (efron) at (5.4,-1) {Efron\\Bootstrap (1979)};
\node[statEvent] (breiman) at (7.5,-1) {Breiman\\CART (1984)};
\node[statEvent] (rumelhart) at (9.5,-1) {Rumelhart\\Autoencoders (1985)};

% Events - AI (above timeline)
\node[aiEvent] (anomaly) at (1.2,1) {Anomaly\\Detection};
\node[aiEvent] (emalg) at (3.3,1) {EM for\\GMMs};
\node[aiEvent] (bagging) at (5.4,1) {Bagging};
\node[aiEvent] (dtrees) at (7.5,1) {Decision\\Trees};
\node[aiEvent] (nlpca) at (9.5,1) {Nonlinear\\PCA};

% Connections
\foreach \s/\a in {parzen/anomaly, demster/emalg, efron/bagging, breiman/dtrees, rumelhart/nlpca} {
    \draw[->, dashed, gray, thin] (\s) -- (\a);
}

% Labels
\node[font=\tiny, orange] at (5.25,-1.8) {Statistical Foundations};
\node[font=\tiny, blue] at (5.25,1.8) {AI Manifestations};

\end{tikzpicture}
\end{figure}

% ====== PART 3: THE MODERN AGE (2000-Present) ======
\begin{figure}[htbp]
\centering
\begin{tikzpicture}[
    timeline/.style={thick, ->, black!60},
    event/.style={rectangle, align=center, font=\tiny, text width=1.6cm, minimum height=0.6cm},
    statEvent/.style={event, fill=orange!20, draw=orange!50, rounded corners=2pt},
    aiEvent/.style={event, fill=blue!20, draw=blue!50, rounded corners=2pt},
    pillarLabel/.style={font=\tiny\bfseries, rotate=90, anchor=center}
]

% Draw timeline
\draw[timeline] (0,0) -- (10.5,0);
\node[anchor=west, font=\tiny] at (10.5,0.3) {Time};

% Title
\node[above, font=\small\bfseries] at (5.25, 2.2) {The Modern Age (2000-Present)};
\node[above, font=\footnotesize] at (5.25, 1.9) {The Explosion into Modern AI};

% Pillar Regions (compact)
\draw[orange, fill=orange!3, very thin] (0.3,-1.5) rectangle (2.2,1.5); 
\node[pillarLabel] at (0.1, 0) {Pillar IV};

\draw[blue, fill=blue!3, very thin] (2.4,-1.5) rectangle (4.3,1.5); 
\node[pillarLabel] at (2.2, 0) {Pillar VI};

\draw[red!70, fill=red!3, very thin] (4.5,-1.5) rectangle (6.8,1.5); 
\node[pillarLabel] at (4.3, 0) {Pillar VII};

\draw[green!60!black, fill=green!3, very thin] (7.0,-1.5) rectangle (8.7,1.5); 
\node[pillarLabel] at (6.8, 0) {Pillar IX};

\draw[violet, fill=violet!3, very thin] (8.9,-1.5) rectangle (10.4,1.5); 
\node[pillarLabel] at (8.7, 0) {Pillar III};

% Events - Statistics (below timeline)
\node[statEvent] (vapnik) at (1.2,-1) {Vapnik\\VC Theory (1995)};
\node[statEvent] (breiman) at (3.3,-1) {Breiman\\Bagging (1996)};
\node[statEvent] (pearl) at (5.6,-1) {Pearl\\SCMs (2000)};
\node[statEvent] (goodfellow) at (7.8,-1) {Goodfellow\\GANs (2014)};
\node[statEvent] (vaswani) at (9.6,-1) {Vaswani\\Transformers (2017)};

% Events - AI (above timeline)
\node[aiEvent] (genbound) at (1.2,1) {Generalization\\Bounds};
\node[aiEvent] (rf) at (3.3,1) {Random\\Forest};
\node[aiEvent] (causalml) at (5.6,1) {Causal\\ML};
\node[aiEvent] (genai) at (7.8,1) {Generative\\AI};
\node[aiEvent] (llm) at (9.6,1) {Large Language\\Models};

% Connections
\foreach \s/\a in {vapnik/genbound, breiman/rf, pearl/causalml, goodfellow/genai, vaswani/llm} {
    \draw[->, dashed, gray, thin] (\s) -- (\a);
}

% Labels
\node[font=\tiny, orange] at (5.25,-1.8) {Statistical Foundations};
\node[font=\tiny, blue] at (5.25,1.8) {AI Manifestations};

\end{tikzpicture}
\end{figure}

\section{The Genealogy of Ideas: A Dependency Graph}
% Here we will place the full dependency graph TikZ picture
% ====== UPDATED GENEALOGY: The Statistical Genealogy of Machine Learning ======
\begin{figure}[htbp]
\centering
\caption{The Statistical Genealogy of Machine Learning Paradigms: A dependency graph tracing the flow of ideas from foundational statistical principles to modern AI methodologies.}
\label{fig:genealogy}
\vspace{0.5cm}
\resizebox{\textwidth}{!}{% Scale to fit
\begin{tikzpicture}[
    concept/.style={rectangle, rounded corners, draw, align=center, text width=2.2cm, minimum height=0.7cm},
    rootConcept/.style={concept, fill=RITorange!40, font=\bfseries\scriptsize},
    midConcept/.style={concept, fill=blue!20, font=\scriptsize},
    leafConcept/.style={concept, fill=green!20, font=\tiny},
    arrow/.style={->, thick, >=stealth}
]

% ====== ROOT NODES (STATISTICAL PRINCIPLES) ======
\node[rootConcept] (inference) at (0,0) {Statistical Inference};
\node[rootConcept] (density) at (0,-1.5) {Density Estimation};
\node[rootConcept] (optim) at (0,-3) {Stochastic Optimization};
\node[rootConcept] (resample) at (0,-4.5) {Resampling Methods};
\node[rootConcept] (replearn) at (0,-6) {Representation Learning};
\node[rootConcept] (explain) at (0,-7.5) {Explainability};
\node[rootConcept] (causal) at (0,-9) {Causal Reasoning};
\node[rootConcept] (kernel) at (0,-10.5) {Kernel Methods};
\node[rootConcept] (ensemble) at (0,-12) {Ensemble Learning};

% ====== MID-LEVEL CONCEPTS ======
% From Inference
\node[midConcept] (mle) at (4,0) {Maximum Likelihood};
\node[midConcept] (bayes) at (4,-0.8) {Bayesian Inference};
% From Density
\node[midConcept] (gmm) at (4,-1.5) {Mixture Models};
\node[midConcept] (kde) at (4,-2.3) {Kernel Density Est.};
% From Optim
\node[midConcept] (sgd) at (4,-3) {Stochastic Grad. Desc.};
\node[midConcept] (em) at (4,-3.8) {EM Algorithm};
% From Resample
\node[midConcept] (boot) at (4,-4.5) {Bootstrap};
\node[midConcept] (xval) at (4,-5.3) {Cross-Validation};
% From RepLearn
\node[midConcept] (pca) at (4,-6) {PCA / Factor Analysis};
\node[midConcept] (mds) at (4,-6.8) {Manifold Learning};
% From Explain
\node[midConcept] (impt) at (4,-7.5) {Feature Importance};
\node[midConcept] (pdp) at (4,-8.3) {Partial Dependence};
% From Causal
\node[midConcept] (scm) at (4,-9) {Structural Models};
\node[midConcept] (iv) at (4,-9.8) {Instrumental Variables};
% From Kernel
\node[midConcept] (svm) at (4,-10.5) {Support Vector Machines};
\node[midConcept] (gaussproc) at (4,-11.3) {Gaussian Processes};
% From Ensemble
\node[midConcept] (bmma) at (4,-12) {Bayesian Model Avg.};
\node[midConcept] (stack) at (4,-12.8) {Stacking};

% ====== LEAF NODES (AI/ML METHODS) ======
% Col 3
\node[leafConcept] (logreg) at (8,0) {Logistic Regression};
\node[leafConcept] (bnn) at (8,-0.8) {Bayesian Neural Nets};
\node[leafConcept] (vaeleaf) at (8,-1.5) {Variational Autoencoders};
\node[leafConcept] (isofor) at (8,-2.3) {Isolation Forest};
\node[leafConcept] (dlopt) at (8,-3) {Deep Learning Optim.};
\node[leafConcept] (gmmleaf) at (8,-3.8) {Training GMMs};
\node[leafConcept] (bag) at (8,-4.5) {Bagging};
\node[leafConcept] (hptune) at (8,-5.3) {Hyperparameter Tuning};
\node[leafConcept] (ae) at (8,-6) {Autoencoders};
\node[leafConcept] (tsne) at (8,-6.8) {t-SNE};
\node[leafConcept] (fi) at (8,-7.5) {Tree-based Importance};
\node[leafConcept] (pdpleaf) at (8,-8.3) {PDP Plots};
\node[leafConcept] (causalmlleaf) at (8,-9) {Causal ML};
\node[leafConcept] (econml) at (8,-9.8) {Econometric ML};
\node[leafConcept] (svmleaf) at (8,-10.5) {SVMs};
\node[leafConcept] (gpleaf) at (8,-11.3) {GPs for ML};
\node[leafConcept] (stackleaf) at (8,-12) {Stacked Ensembles};
\node[leafConcept] (nnet) at (8,-12.8) {Neural Net Ensembles};

% Col 4
\node[leafConcept] (transform) at (12,-3) {Transformers (NTK)};
\node[leafConcept] (gan) at (12,-1.5) {Generative Adversarial Nets};
\node[leafConcept] (xboost) at (12,-4.5) {Gradient Boosting (XGBoost)};
\node[leafConcept] (shap) at (12,-7.5) {SHAP Values};
\node[leafConcept] (llm) at (12,-9) {Causal LLMs};

% ====== DRAW ARROWS ======
% Connect Roots to Mid-Level
\foreach \source/\target in {inference/mle, inference/bayes, density/gmm, density/kde, optim/sgd, optim/em, resample/boot, resample/xval, replearn/pca, replearn/mds, explain/impt, explain/pdp, causal/scm, causal/iv, kernel/svm, kernel/gaussproc, ensemble/bmma, ensemble/stack} {
    \draw[arrow] (\source) -- (\target);
}
% Connect Mid-Level to Leaves
\foreach \source/\target in {mle/logreg, bayes/bnn, gmm/vaeleaf, kde/isofor, sgd/dlopt, em/gmmleaf, boot/bag, xval/hptune, pca/ae, mds/tsne, impt/fi, pdp/pdpleaf, scm/causalmlleaf, iv/econml, svm/svmleaf, gaussproc/gpleaf, bmma/stackleaf, stack/nnet} {
    \draw[arrow] (\target) -- (\source); % Reversed for layout
}
% Connect Special Cross-Pillar Dependencies
\draw[arrow, dashed, bend left=10] (kde) to (gan);
\draw[arrow, dashed, bend left=10] (gmm) to (gan);
\draw[arrow, dashed, bend left=10] (sgd) to (transform);
\draw[arrow, dashed, bend left=10] (kernel) to (transform);
\draw[arrow, dashed, bend left=10] (boot) to (xboost);
\draw[arrow, dashed, bend left=10] (optim) to (xboost);
\draw[arrow, dashed, bend left=10] (impt) to (shap);
\draw[arrow, dashed, bend left=10] (ensemble) to (shap);
\draw[arrow, dashed, bend left=10] (scm) to (llm);
\draw[arrow, dashed, bend left=10] (replearn) to (llm);

\end{tikzpicture}
} % End resizebox
\end{figure}
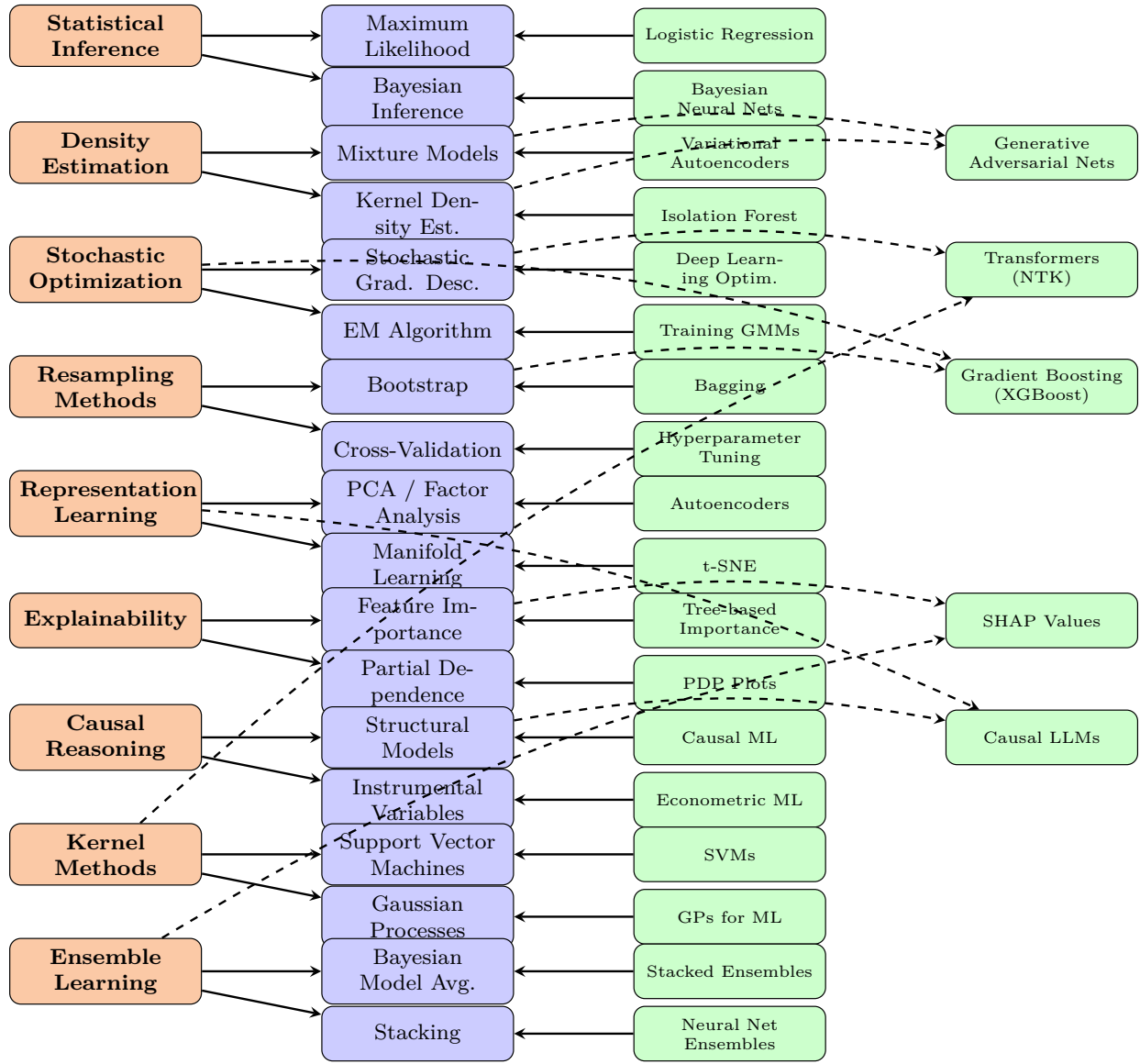
\end{appendix}

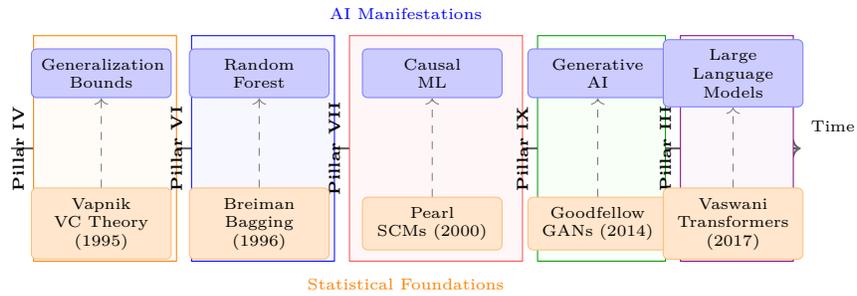
\end{document}